\documentclass{cimart}

\title[The PORTSEA and a few personal scientific achievements]{The PORTSEA (Portuguese School of Extremes and Applications) and a few personal scientific achievements}

\author{Maria Ivette Gomes}

\authorinfo{Centre of Statistics and its Applications, University of Lisbon, Portugal}{migomes@ciencias.ulisboa.pt}

\abstract{%
The {\it Portuguese School of Extremes and Applications} (PORTSEA) is nowadays well recognized by the international scientific community, and in my opinion, the organization of a NATO {\it Advanced Study Institute} on {\it Statistical Extremes and Applications}, which took place at Vimeiro in the summer of 1983, was  a landmark for the international recognition of the group and the launching of  the PORTSEA. 
The dynamic of publication has been very high and the topics under investigation in the area of {\it Extremes} have been quite diverse. In this article, attention will be paid essentially to some of the scientific achievements of the author in this field, but apart from a large group, where the author is  included,  working in the area of {\it Parametric, Semi-parametric and Non-parametric  Estimation of Parameters of Rare Events}, the PORTSEA has strong groups in {\it Univariate, Multivariate, Multidimensional, Spatial Extremes and Applications to Dynamical Systems, Environment, Finance} and {\it Insurance}, among others.
We thus think that the dynamism of the Group will provide a healthy growing of the field, with a high international recognition of  {\it Extremes} in Portugal, a country of  ‘{\it good extremists}’ in an extreme of Europe.}



\VOLUME{32}
\ISSUE{3}
\firstpage{329}
\DOI{https://doi.org/10.46298/cm.13109}

\begin{document}

\tableofcontents

\section{A brief introduction}
I would dare to say that the PORTSEA ({\it Portuguese School of Extremes and Applications}) recognition is mainly due to the scientific work of Tiago de Oliveira in the area. But I also need to mention the research developed by Feridun Turkman and the author, while working for PhD in Sheffield, {\it United Kingdom} (UK), and the organization of a NATO {\it Advanced Study Institute} (ASI) on {\it Statistical Extremes and Applications} (SEA), which took place at Vimeiro in the summer of 1983.
Indeed,  the organization of this 1983 NATO ASI (SEA 1983) was  a landmark for the international recognition of the group and the launching of  the PORTSEA, but other relevant items were also crucial for such a recognition, as can be seen in \mbox{{\bf Section \ref{Sec-PORTSEA}}}. A few details on the main topics developed along my  research carreer  in the area of {\it Extremes} will be provided in {\bf Section \ref{Sec-Personal}}. I shall finally draw  a few  overall comments in \mbox{{\bf Section \ref{Sec-Final}}}.

\section{The building  of PORTSEA\label{Sec-PORTSEA}}

\subsection{Prior to its existence\label{Sec-2.1-Prior}}
After getting an undergraduate degree  in {\it Pure Mathematics} ({\it Algebra}), in 1970, at ``{\it Facul\-dade de Ci\^encias da Universidade de Lisboa}'' (FCUL), I took the decision to go on with research in the area of {\it Probabi\-lity, Statistics and Stochastic Processes}, becoming then  a member of  the {\it Statistics} group,  at ``{\it Departamento de Matem\'atica}'', FCUL, under the supervision of  Professor Jos\'e Tiago da Fonseca Oliveira (Tiago de Oliveira, from now on), already a  prominent international reference in the area  of  SEA, with several articles on this topic  since 1959, as can be seen in the preprint \cite{MIG-2021}, in English, and in the recent preliminary version of a book, \cite{MIG-2023a}, in Portuguese, written by the author. 
\vspace{.25pc}

During the period I worked in Portugal as a research assistant (1970--1975),  first at ``{\it Centro de Matem\'atica Aplicada}'', and next at ``{\it Centro de Estat\'istica e Aplica\c{c}\~{o}es da Universidade de Lisboa}'' (CEAUL), a Research Centre founded in 1975 by  Tiago de Oliveira, and despite  the fact that I had not so far worked in the field of   {\it Extreme Value Theory} (EVT),  I was well acquainted with the potentialities and beauty of  EVT, first due to the reading of Gnedenko’s book \cite{Gnedenko-1969} and also due to some articles written by  Tiago de Oliveira. 
\vspace{.25pc}

The investment policy inspired by Veiga Sim\~ao, a true reformer of Higher Education in Portugal,   opened up great prospects for researchers considered promising, who were sent to major cultural centres abroad. 
My husband, Dinis Pestana, and I  got a Calouste Gulbenkian fellowship and were  accepted at the University of Sheffield,  UK. 
Working also for PhD\ in Sheffield, in a related area, but under the supervision of Morris Walker, was Kamil Feridun Turkman, now Full Professor (already retired) of the ``{\it Departamento de Estat\'{\i}stica e Investiga\c{c}\~ao Operacional}'' (DEIO), and Ant\'onia Amaral (now Ant\'onia Amaral Turkman), a colleague and friend  of mine at FCUL,  who began working for PhD by the end of 1977, also  in Sheffield, but in the area of {\it Bayesian Statistics},  under the supervision of Ian Dunsmore.
\vspace{.25pc}

SEA was then (and it is still now \dots) considered as a quite relevant area in the field of {\it Statistics}, with a lot of topics to be  explored. Indeed, in the first decades of the twentieth century, under the powerful influence of Paul L\'evy, {\it Probability Theory} was mainly concerned with generalizations of the {\it Central Limit Theorem} (CLT), related to the asymptotic behaviour of sums---a problem of great importance, since averages, variances and many other relevant statistics are simple sum functions.
But  Maurice Fr\'echet \cite{Frechet-1927}, in 1927,   
had   the interesting idea of using an analogue of the L\'evy stability equation for sums, just noticing that the maximum of maxima is still a maximum, i.e.\ the  maximum of the $m\times n$ values, $X_1,X_2,\dots,X_{m\times n}$, is also the maximum of the  $n$ maxima values of $X_{(i-1)\times m+1},\dots,X_{i\times m}, \ 1\leq i\leq n$. 
He then just replaced powers of characteristic functions by  powers of distribution functions, that is, he treated a problem analogous to that of sums, but for maxima of {\it independent, identically distributed} (IID) {\it random variables} (RVs). 
He  thus came to the functional stability equation for maxima, which enabled the definition of a {\it max-stable} (MS) distribution, i.e.\   a {\it cumulative distribution function} (CDF)  $G$ for which there exist sequences of real constants,   $A_k>0$ and $B_k\in\mathbb{R}$ such that
\begin{equation}
 G^k(A_k x+ B_k)= G(x)\quad \mbox{for all } k\geq 1.
\label{Eq-MS-equation}
\end{equation}
He has thus `invented’ the first law of extremes, rightly called Fr\'echet  CDF, with a  functional form of the type
$$
\Phi_\alpha(x)= \exp(-x^{-\alpha}), \ x\geq 0\quad (\alpha>0),
$$
one of the solutions of \eqref{Eq-MS-equation}.
 Almost at the same time, Fisher and Tippett \cite{F+T-1928}, in 1928,   discovered the three types of solutions to which the max-stability equation, in \eqref{Eq-MS-equation}, can lead,
\begin{equation*}
\begin{array}{ll}
\mbox{Type I (Gumbel)}: & \Lambda(x)=\exp(-\exp(-x)),\ x\in \mathbb{R},\\
\mbox{Type II (Fr\'echet)}: &  \Phi_\alpha(x)=\exp(-x^{-\alpha}), \ x\geq 0\quad (\alpha>0), \\
\mbox{Type III (max-Weibull)}: &  \Psi_\alpha(x)=\exp(-(-x)^{\alpha}), \ x\leq 0\quad (\alpha>0).
\end{array}
\label{Tipos1-3}
\end{equation*}
Meanwhile, von Mises  \cite{Mises-1936}, in 1936,  proposed an expression encompassing these three laws, \begin{equation}
G(x)\equiv G_\xi(x):=\left\{
\begin{array}{lll}
\exp\left(-(1+\xi x)^{-1/\xi}\right),\ 1+\xi x>0, & \mbox{if} & \xi\not=0,\\
\exp(-\exp(-x)),\ x\in\mathbb{R},  & \mbox{if} & \xi=0, 
\end{array}
\right.
\label{Eq-EVI-xi}
\end{equation}
the so-called {\it general extreme value} (GEV) CDF, worked out from a statistical point of view by Jenkinson  \cite{Jenkinson-1955}, and sometimes also  called the {\it von Mises-Jenkinson} CDF.
Indeed, with $\xi=0$, $\xi= {1}/{\alpha}>0$ and $\xi=- 1/{\alpha}<0$, respectively, we have  $\Lambda(x)=G_0(x)$, $\Phi_\alpha(x)=G_{1/\alpha}(\alpha(1-x))$  and   $\Psi_\alpha(x)=G_{-1/\alpha}(\alpha(x+1))$.
The generalized shape parameter $\xi$ ($\in\mathbb{R}$) is the so-called {\it extreme value index} (EVI), the main parameter in EVT.
Such a result was initially formalized by Gnedenko \cite{Gnedenko-1943}, in 1943, used by  Gumbel \cite{Gumbel-1958}, in 1958,  for applications of EVT, essentially in engineering and hydrology, and later fully formalized by de Haan \cite{Haan-1970}, in 1970,  in what could be called the {\it Fisher-Tippett-Gnedenko-de Haan Theorem}, but is usually known as the {\it Fisher-Tippett-Gnedenko Theorem} or the {\it Extremal Types Theorem} (ETT).
Currently (for more details, see the reasonably recent overview by Gomes and Guillou  \cite{G+G-2015}), these results are indeed unified in a general theory, which recognizes  that the  ETT and other ETT generalizations  (see \cite{M+P+G-2015, B+G+M+P+P-2023}, among others) are  references for the study of extreme {\it order statistics} (OSs), while the CLT has to do with sums and central OSs. 
\vspace{.25pc}

Statistical EVT  had indeed a strong   development under the vigorous impulse of  Emil Julius Gumbel, in the late sixties. 
And  the {\it School of Extremes} (and {\it Risk Evaluation}) in Portugal, or the PORTSEA, is nowadays well recognized by the international scientific community, being  such a recognition  mainly due to the scientific work of Tiago de Oliveira, Effective Member of the ``{\it Academia das Ci\^{e}ncias de Lisboa}’’ (ACL)  from 1985 until his premature death in 1992. Possibly due to his political ideas, Tiago de Oliveira was a good friend of Emil Gumbel, who deeply influenced his change from {\it Algebra} to {\it Statistics of Extremes} (see, \cite{TiagoO-1959}, Tiago’s  first article in the area, as far as I know). 
\vspace{.25pc}

But I also need to mention the research developed by myself and by Kamil Feridun Turkman in the field, while working for PhD in Sheffield, United Kingdom,  in the late seventies---early eighties. 
I  received  the PhD degree by the end of 1978 (see \cite{MIG-1978}), under the supervision of Clive Anderson, an eminent scientist in the field of {\it Extremes}. Dinis Pestana, my husband, also got his PhD in Sheffield, in 1978, in topics related to sums of RVs, under the supervision of Damodar N. Shanbhag. We both came then  back to  FCUL by the end of 1978, and collaborated actively in the foundation in 1980 (November, 28) of the ``{\it Sociedade Portuguesa de Estat\'{\i}stica e Investiga\c{c}\~ao Ope\-racional}'' (SPEIO), with Tiago de Oliveira as the first President. SPEIO was profoundly restructured in 1991, with the current designation, ``{\it Sociedade Portuguesa de Estat\'{\i}stica}'' (SPE), and I was the first President of SPE, a Society with a very high impact in the development of {\it Statistics} in Portugal.  
\vspace{.25pc}

In 1981,  and after getting his PhD  (see \cite{Turkman-1980}), also  in Sheffield, in the area of {\it Extremes in Stochastic Processes} and  under the supervision of Morris Walker, Feridun Turkman  joined us at FCUL  and at CEAUL,  the main pole of development of  {\it Extremes} in  Portugal.   Ant\'{o}nia Amaral-Turkman, Dinis Pestana, Feridun Turkman and the author, Ivette Gomes,  together with Cristina Sernadas (also from {\it Statistics}) and colleagues from {\it Operations Research} (J. Dias Coelho) and {\it Computing} (Am\'{\i}lcar Sernadas), all young people who had also got their PhDs abroad,  worked   hard, jointly with Tiago de Oliveira,   in the foundation, in 1981, of the ``{\it Departamento de Estat\'{\i}stica, Investiga\c{c}\~ao Operacional e Computa\c{c}\~ao}''  (DEIOC), now DEIO, with the first degrees in the area of {\it Statistics}, in Portugal, one in {\it Probability and Statistics} and another one in {\it Statistics and Operations Research}.
\vspace{.25pc}

In the mid of 1981, after the formation of  DEIOC,   Feridun and I,  jointly with Tiago de Oliveira, have proposed the organization of the  NATO ASI on SEA, which took place at Vimeiro in the summer of 1983 (SEA 1983). 
As already mentioned  here, and also in \cite{MIG-2005, MIG-2007a, MIG-2023b},  articles written in Portuguese,   the organization of the 1983 NATO ASI was  indeed  a landmark for the international recognition of the `{\it School of Extremes}' in Portugal.
And although they have not usually been considered as  elements of this group, due to the fact that their main topics of research are not in the field of {\it Extremes}, I am sure that  Ant\'onia Amaral Turkman  and Dinis Pestana  have also played a significant role in the construction of the  group, even co-authoring several  relevant articles in this area, and I consider that they are indeed members of the PORTSEA, and `shadow organizers’ of SEA 1983.

\subsection{SEA 1983  and the launching of PORTSEA\label{Sec-2.2-Launching}}
\begin{figure}[ht]
    \centering
    \includegraphics[bb=0 0 379 261]{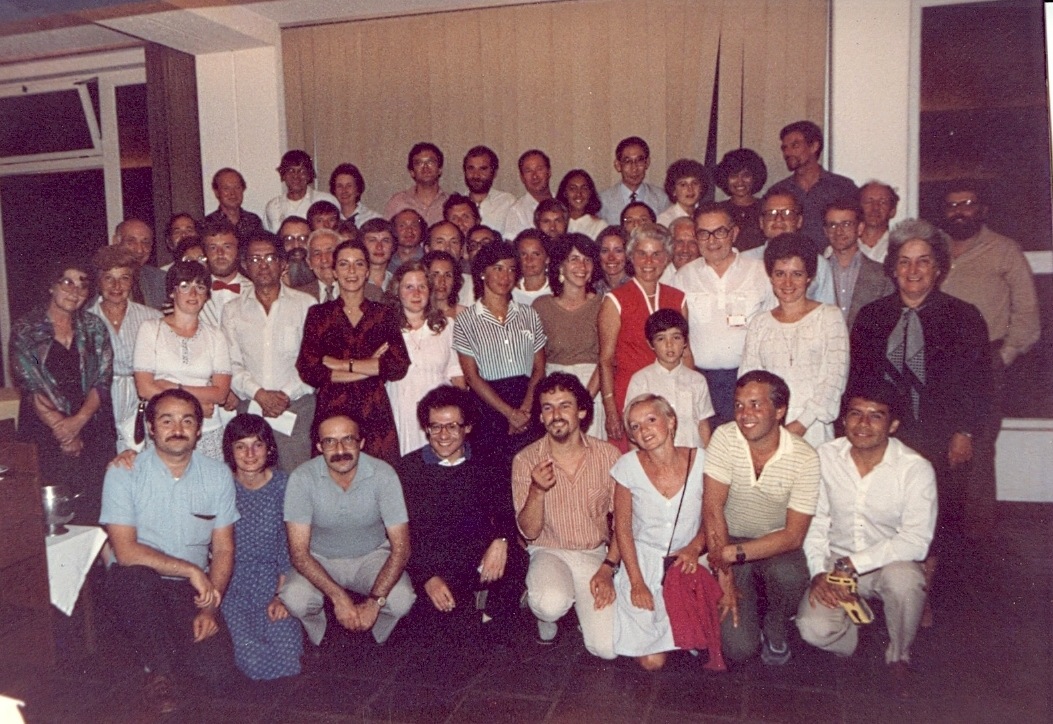}
    \caption{Photo of NATO ASI participants---Vimeiro (SEA 1983)}
    \label{ASI-participants}
\end{figure}

The SEA 1983 NATO ASI was held in Vimeiro,  from 31st of August    until  September 14, and had the participation of prominent researchers in the area, with  some of them  present in the  photo provided   in Figure \ref{ASI-participants}.

Among those participants,  I mention the invited foreign speakers,  Clive Anderson
(Sheffield University, UK),  Paul Deheuvels (Universit\'e Paris VI, France),   Benjamin Epstein (Technion, Israel), Janos Galambos (Temple University, USA), Arne Frans\'{e}n (National Defence Research Institute, Sweden), Laurens de Haan (Erasmus University of Rotterdam, The Netherlands),  Leon Herbach (Polytechnic Institute of New York,  USA), Michael Hasofer (University of New South Wales, Australia), Ross
Leadbetter (University of North Carolina, USA), Georg Lindgren (University of Lund, Sweden), Nancy Mann (Department of Biomathematics, UCLA, USA), B. Marcus (Texas A\&M University, USA),  Yashaswini Mittal (Virginia Polytechnic Institute and State University, USA), James Pickands III (University of Pennsylvania, USA), Sid Resnick (Colorado State University, USA), Holger Rootz\'en (University of Copenhagen and UNC, Chapel Hill, USA), Gerhart  Schu\"{e}ller (Institut fur Mechanik, Universitat Innsbruck, Austria), Masaaki Sibuya (Keio University, Japan), R. Sneyers (Royal Meteorological Institute, Brussels), Jef Teugels (Katholieke Universiteit Leuven, Belgic),  Ishay Weissman (Technion, Israel), Vujica Yevjevich (International Water Resources Institute, George Washington University, USA). 
\vspace{.25pc}

Also, some of the students of the first DEIOC MSc\  course on {\it Probability and Statistics},   Teresa Alpuim,  Em\'{i}lia Athayde,   Isabel Bar\~ao and  F\'atima Migu\'ens, together with Manuela Neves and Fernando Rosado,  both PhD students of Tiago de Oliveira, and two PhD students of Dinis Pestana, Eug\'{e}nia Gra\c{c}a Martins and Helena Igl\'{e}sias Pereira, were young participants of this NATO ASI. 
And  in the list of authors of `{\it contributed papers}'  we can find names of giants in the area of {\it Extreme Value Analysis} (EVA),  like Richard Davis, who was the 2023 `{\it Pedro Nunes Lecturer}’, Anthony Davison, J\"{u}rg H\"{u}sler,  Rolf Reiss, Richard Smith \dots 
\vspace{.25pc}

As mentioned above, this NATO ASI on SEA is currently recognized as a milestone in the area of EVA.  
And repeating again what I said before in several occasions (see, for instance, the interviews in \cite{B+J+AP+V-2015,FA+DC-2015,ACMF+JMF-2018,Lourenco-2022}, it was indeed true that when Richard Davis, one of the organizers of  EVA 2009 (The {\it Sixth International Conference on Extreme Value Analysis}), which was held at Fort-Collins, Colorado,  USA,   spoke about Vimeiro's meeting as  EVA--0, and when  I read at  EVA 2013 website:
`{\it It has been $30$ years since the so-called zero-th {\rm EVA} conference took place in $1983$ in Vimeiro, a small town near the beach in Portugal}' \dots 
I indeed felt some `Nostalgia'  \dots 
\vspace{1pc}

 \noindent\begin{minipage}{0.3\textwidth}
 \begin{itemize}
 \item []
\includegraphics[width=\linewidth, bb=0 0 545 825]{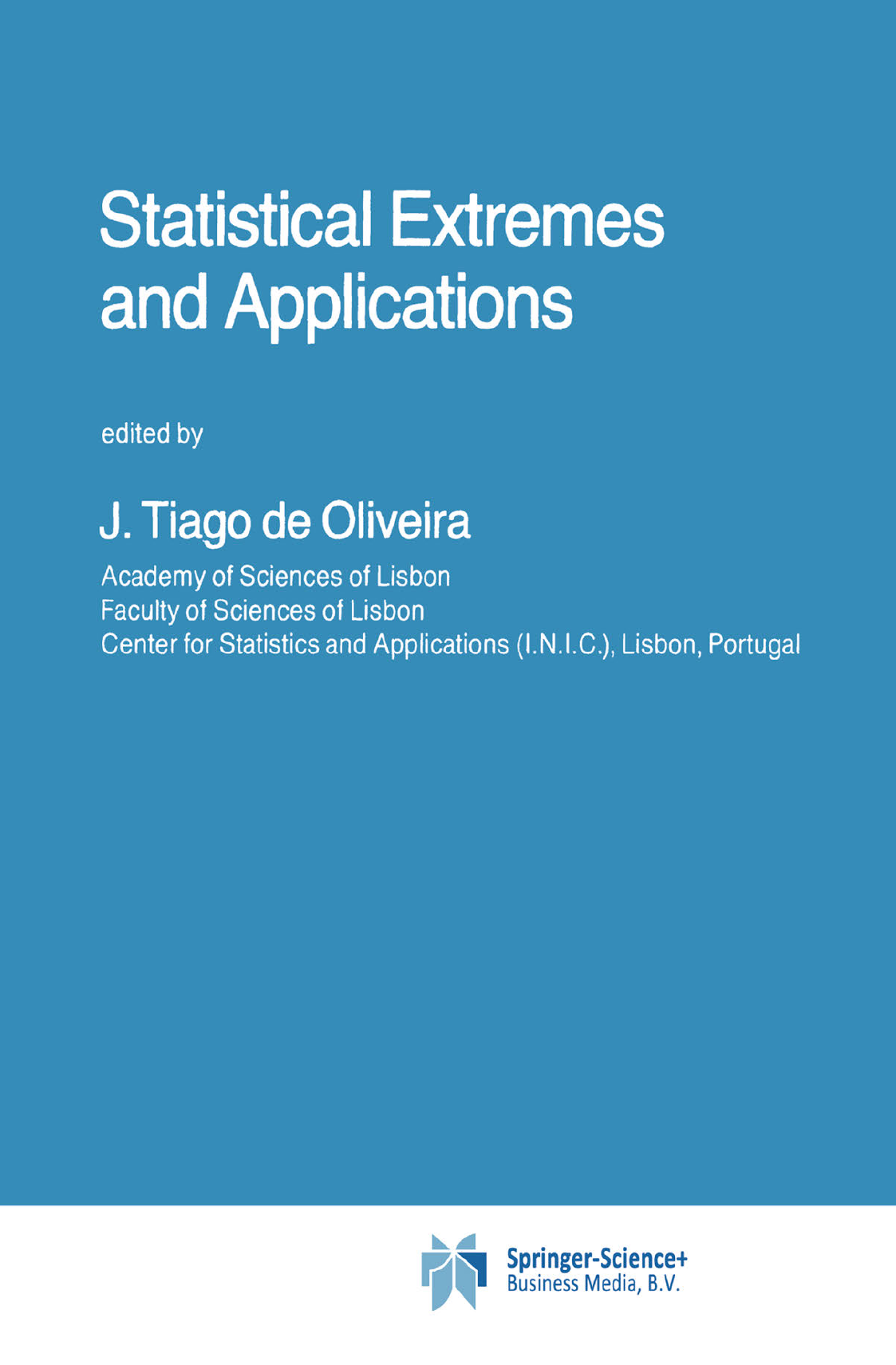}
\end{itemize}
\end{minipage}%
\hfill%
\begin{minipage}{0.6\textwidth}
\begin{itemize}
\item  [] In the Preface of the book \cite{TiagoO-1984} associated with this NATO ASI, edited by Tiago de Oliveira  and dedicated to the memory of Emil Julius Gumbel, one of the pioneers in {\it Statistics of Extremes}, 
and a scientist with whom Tiago de Oliveira collaborated in the sixties,   at Columbia University, 
one can find articles written by the aforementioned  prominent researchers. 
\end{itemize}
\end{minipage}
 \vspace{1pc}

In such a Preface  we can read: ‘… {\it the narrow and shallow stream} (of extremes) {\it gained momentum and is now a huge river, enlarging at every moment and flooding the margins}’.  And Tiago de Oliveira ends the Preface with thanks to the members of the recently formed DEIOC/FCUL, now DEIO/FCUL, saying: ‘\dots {\it it is a very good group that crossed the desert during the organization time and continues to work on}\dots' 
\vspace{.25pc}

The urge to publish was then reduced. Two of the most relevant results contained in my 1978 thesis,
 the derivation of the joint distribution of upper OSs and their concomitants, or induced OSs (see  {\bf  \ref{Sec-upperOSs}}), 
 and the study of rates of convergence and penultimate or pre-asymptotic  behaviour of sequences of extremes (see {\bf  \ref{Sec-Velocidade}}),
were published only in 1981  \cite{MIG-1981a} and in 1984 \cite{MIG-1984c}, respectively. 
\vspace{.25pc}

In that distant 1983, the community of  `extremists' was just emerging. There were {\bf 15}  days of intense exchange of ideas, which renewed my enormous enthusiasm for EVT. 
And EVT has developed rapidly in recent decades due to its importance in the assessment of catastrophic risks in the most diverse  human activities, among which I mention, {\it Dynamical Systems}, {\it Economy}, {\it Finance}, {\it Health}, {\it Industry}, {\it Insurance} and {\it Population Dynamics}. EVT is essential for the construction of large structures in which it is necessary to assess levels of exceedance, for example of wind speeds or of  river flows during floods. And it is one of the instruments of research in  {\it Climatology}, {\it Energy}, {\it Environment},  {\it Hydrology} ---in short, EVT has invaded almost all fields of science and technology related to a collective survival,  where  parameters of rare events are extremely relevant. 
That is why we welcome the important international impact of Portuguese ‘{\it extremism}’, whose success is bound to be increasingly visible. 
\vspace{.25pc}

I now dare to say again (see, \cite{MIG-2005, MIG-2007a, MIG-2013b, MIG-2017a, MIG-2023a, MIG-2023b}) that the organization of the 1983 NATO ASI (or SEA 1983), with two Wednesday's afternoons and two whole weekends, full of social program, under my responsibility, despite of a bit  `traumatizing', in such a way that only after 1999 did I advance with the organization of conferences in Portugal,  was  indeed  a landmark for the international recognition of the group and the launching of the  PORTSEA, with an active life of more than {\bf 40}  years, if we count it only after September 1983. 
\vspace{.25pc}

\subsection{Other founders of the PORTSEA\label{Sec-2.3-Other}}
With some work in the area of {\it Extremes}, although slight collateral to the subject of their  PhD, I also   mention  Eug\'enia Gra\c{c}a Martins,  with a PhD thesis discussed in 1983,  Helena Igl\'esias Pereira,  who got her PhD degree at FCUL, in 1985, both supervised by Dinis Pestana, and Fernando Rosado, who got his PhD in 1984, under the supervision of Tiago de Oliveira.  
 In the area of {\it Risk and Ruin Theory}, which can potentially be partially included in the PORTSEA,  something `{\it I hope for the future}', I mention Lourdes Centeno and Alfredo Eg\'{i}dio dos Reis, who got their PhD degrees  in 1985 \cite{Centeno-1985} and 1994 \cite{ReisE-1994} respectively, at Heriot-Watt University (UK).
 \vspace{.25pc}
  
  In the meantime,  Margarida Brito (University of Porto), who  obtained her PhD degree in 1987  in the area of {\it Extremes} at the University of Paris VI  \cite{Brito-1987}, under the supervision of Paul Deheuvels, came back to Portugal, being now an Associate Professor at the University of Porto. 
 \vspace{.25pc}

From the beginning of the 1980s, the investment policy  inspired by  Veiga Sim\~{a}o  began to bear fruit, in the sense that a few  groups with a reasonable  critical mass were created  in Portugal. 
And, together with a few PhDs got in the best  centres out of Portugal, those groups allowed  the beginning of a `{\it banal}'  supervision of PhD's in Portugal, together with PhDs obtained  in the best foreign centres.  
I next sequentially mention those first  PhD thesis:

\begin{itemize}
\vspace{-.5pc}
\item [---] The first student to get a PhD\ degree in Portugal fully  in the area of  {\it Extremes} was Teresa Alpuim, now Full Professor at DEIO. She got her PhD\ degree in 1989, under my supervision, at the University of Lisbon \cite{Alpuim-1989}. Her thesis gave rise to  seminal papers in the field of {\it Extremes for Dependent Sequences}, as can be seen in \cite{MIG-2023a}. 
\vspace{-.5pc}
\item  [---] Almost simultaneously, but already in 1990, M. Manuela Neves, now Retired Full Professor at the ``{\it Instituto Superior de Agronomia}’’ (ISA), ``{\it Universidade T\'{e}cnica de Lisboa}’’ (UTL), now University of Lisbon, defended her PhD\ thesis \cite{NevesMM-1990},   at   ``{\it Universidade Nova de Lisboa}’’ (UNL),    under the supervision of  Tiago de Oliveira. 
She was thus the first PhD student of Tiago de Oliveira in the field of {\it Extremes}.
\vspace{-.5pc}
\item   [---] Also pioneers in this area are  my second and third PhD\ students,   Lu\'{\i}sa  Canto e Castro \cite{Canto-1992} and  Isabel Fraga Alves \cite{FragaA-1992}, who defended their PhD\ theses in 1992.  
\vspace{-.5pc}
\item   [---] Fernanda Oliveira defended her PhD thesis \cite{OliveiraMF-1992}  also in   1992, under the supervision of Feridun Turkman. 
\vspace{-.5pc}
\item   [---] Although in a collaborative way,  I also consider  Nuno Crato, who got his PhD degree at the University of Delaware, USA  \cite{Crato-1992}, under the guidance of Howard Taylor, 
 as a pioneering name for the development of what I dare today to call the PORTSEA.

\vspace{-.5pc}
\item   [---] Professor Tiago de Oliveira only quite late decided for the supervision of PhD\ students in his most relevant area of research, and unfortunately, due to his premature death in 1992, when he was 63 years old, he has only seen the discussion of  Manuela Neves' PhD\ thesis.  
 Two other PhD\ students of  Tiago Oliveira, in the area of {\it Extremes},  and members of DEIO/FCUL,  Isabel Bar\~ao and   Teresa Themido Pereira, have finished their PhD\ thesis,  already under my supervision  \cite{Barao-1993,ThemidoPereira-1994}. 
 \vspace{-,5pc}
 \item  [---] Also in 1994,  Em\'{i}lia Athayde \cite{Athayde-1994}    at FCUL, and Helena Ferreira \cite{FerreiraH-1994}   at the University of Coimbra, got PhD degrees  under my supervision.  
 
 \end{itemize}

This was the beginning of the Portuguese `{\it extremism}'. First under the direct guidance of the aforementioned researchers,  and next by the  scientific `grandchildren'  of  initial members, was created  the PORTSEA, recognized internationally, whose members are spread  throughout all Portuguese universities,  and even around the world. 
 For some, the passage through the universe of {\it Extremes} was fleeting, directing their interests to other areas, but most of the researchers who got their PhD in the area continued to publish firmly in the wide range of EVA, and many others whose usual research activities are developed in other areas have occasionally produced valuable work in the field of {\it Extremes}.
\vspace{.25pc}

\subsection{The growth of PORTSEA---Further PhD Thesis and Habilitation Degrees\label{Sec-2.4-Growth}}

From the beginning of  1994 and until the end of 2003, in a period of 10 years, the PhDs of Portuguese researchers in the area of  {\it Extremes} have followed at a quite interesting rhythm. And apart from the {\bf 3} aforementioned PhD thesis written by Teresa Themido Pereira,   Em\'{\i}lia Athayde and Helena Ferreira, in 1994,  {\bf 18} additional  PhD thesis were mentioned in \cite{MIG-2021, MIG-2023a}. 
\vspace{.25pc}

It is indeed sensible to refer that Laurens de Haan, one of the giants in the area of {\it Extremes}, and author of a highly cited  PhD thesis \cite{Haan-1970},  has regularly visited Lisbon since 1997, and this has led to the development of joint research work with several members of CEAUL.  He came  to Portugal in 1999, becoming  then a member of CEAUL and of the PORTSEA.  
On the grounds of the strong cooperation developed between Laurens de Haan and members of DEIO/FCUL, and even more generally his cooperation with members of the Portuguese statistical community, DEIO has proposed a title of University of Lisbon  ``{\it Doutor Honoris Causa}’’  to Laurens de Haan. He has accepted such a distinction,  and the   title was awarded in 2000.  
And in 2013  another giant in the field, Ross Leadbetter, 
who died in March 2022, and received several homages of members of the PORTSEA   (see \cite{MIG-2022c, MIG-2022c, MIG-2022d, G+T-2022}), 
has honored the University of Lisbon by accepting the same distinction, since no doubt, and just as I wrote at {\it Info-Ci\^{e}ncias Digital} (see \cite{MIG-2013a}),  when the university honors researchers of this importance it is  also honored. The group has thus two of the  {\it University of Lisbon Honoris Causa Doctors}.
\vspace{.25pc}

After 2004 and up to the end of 2018, in a period of 15 years, the growth rate decreased slightly. Anyway, and being sure that a few PhD thesis are missing,   {\bf 29} additional PhD thesis in the field could be counted  in \cite{MIG-2021},  most of them supervised by members of our PORTSEA.
\vspace{.25pc}

 The number of Habilitation Degrees has not been as high as expected. Apart from the {\bf 5}  initial members of the group, I could  count only {\bf 14} additional  people with an {\it Habilitation Degree} (Lourdes Centeno, 1993; Teresa Alpuim, 2002; Nuno Crato, Alfredo Eg\'idio dos Reis, 2002; Manuela Neves, 2003; Isabel Fraga Alves, 2003; Helena Ferreira, 2004; Luisa Pereira, 2008;  Manuel Scotto, 2012;  Jorge Milhazes de Freitas; 2014; Ana Cristina Moreira  Freitas, 2014; F\'atima Brilhante, 2015;  Ana Ferreira, 2016; Miguel De Carvalho, 2018). But I may be  forgetting somebody, and I believe that a few members of the PORTSEA will soon apply for such a degree.
\vspace{.25pc}

\subsection{Conferences' organization under the PORTSEA umbrella\label{Sec-2.5-Conferences}}

After the organization of the 1983 NATO ASI on SEA,  Feridun Turkman (jointly with Vic Barnett, University of Sheffield) organized in 1993, a SPRUCE meeting on {\it Statistics for the Environment},  in Lisbon,  where {\it Extremes} played a very  important role (see \cite{B+T-1993}). Other SPRUCE meetings have been co-organized by Feridun, out of Portugal. 
But  we can say that after  a wide interregnum of about 15 years, these last two decades have been fruitful in the organization of international conferences in Portugal in the area of {\it Extremes}, with the inclusion of the area of  {\it Risk Analysis}, where tails are also quite relevant. The {\bf 5} organizations of international conferences referred to in \cite{MIG-2007a}  are   {\bf 16}, in \cite{MIG-2023a}, including the following ones, which I think sensible to recall again:

\begin{enumerate}

\item {\it Workshop on Statistical Modelling---Extreme Values and Additive Laws}, Estoril, October 2--7, 1999. This workshop ran under the sponsorship of  CEAUL and the ``{\it Funda\c{c}\~{a}o para a Ci\^{e}ncia e a  Tecnologia}’’ (FCT) project `MODEST---{\it Statistical Modeling}', a project developed in the interaction of two sub-projects: the subproject `MECAES---{\it Stochastic Models in Environment, Ecology and Health Sciences}', leaded by Kamil Feridun Turkman,  and the subproject `VELA---{\it Extreme Values and Additive Laws}', leaded by M. Ivette Gomes (1997-2000). For details, on all accepted conference papers, see \cite{G+P+CC+FA+M-1999}.

\item EMS 2001: 23rd {\it European Meeting of Statisticians}, Funchal, Madeira,  August, 13--18,   2001. The EMS 2001 was organized under the auspices of the {\it European Regional Committee of the Bernoulli Society}, in a consortium with the  University of Lisbon, the  ``{\it Universidade da Madeira}’’ (UMa) and the  ``{\it Instituto Nacional de Estat\'istica}’’ (INE), and had more than  500 participants.
There was then a strong cooperation among  CEAUL,   INE and  UMa, and a high recognition of the area of   {\it Extremes}.  I have organized an invited session, entitled  “{\it Statistics of Extremes}”, where I was asked to be one of the speakers (see \cite{MIG-2001}). The other invited speakers were Jan Beirlant, Holger Rootz\'{e}n and  Ross Leadbetter. 
After a light revision, the submitted abstracts were published in three volumes of  the  ``{\it Revista de Estat\'istica}’’, edited by   Davison, Ferreira da Cunha,  Fraga Alves \& D Pestana (see \cite{D+P+FC+FA-2001}).
And the EMS 2001 was the launching  for the journal {\it Revstat---Statistical Journal}. Indeed, INE was editing the above-mentioned  national journal, ``{\it Revista de Estat\'istica}’’, and had decided to try to transform it into an international journal at the beginning of 2001. I was then invited  by Ferreira da Cunha, Editor of ``{\it Revista de Estat\'istica}’’, to be Editor-in-Chief of an international continuation of this national journal. I took advantage of the high number of renowned statisticians present in Funchal, to begin the construction of the first possible set of {\it Associate Editors}. Among them, I mention names like David Cox, Isaac Melijson and Jef Teugels. After talking to them, I recognized that it would indeed be possible to form a solid editorial board and accepted the invitation. And I had colleagues who from the very beginning were very excited about the launch of {\it Revstat}, and who helped a lot. Some of them, like Ant\'onia Amaral Turkman and Dinis Pestana, were from Portugal. I was not alone, fortunately, but we had a lot of work to launch the first volume of {\it Revstat---Statistical Journal}, which only came out in 2003.

\item  EVA 2004---{\it Third International Symposium on Extreme Value Analysis: Theory and Practice}, Aveiro, July 19--23, 2004. After Gothenburg  (1998) and Leuven (2001), EVA 2004 was held in Aveiro, in a certain sense as a recognition for what we already had done in the field (see \cite{H+G+R+S-2004}). The ‘{\it extremists}’ and PORTSEA members Andreia Hall and Manuel Scotto played then  a very important role.

\item ISI WSC 2007: 56th {\it Session of the International Statistical Institute World Statistical Conference}. This large event, with more than 2000 participants, was held in Lisbon, Portugal, 22-29 August. I was the Chair of the Local Program Committee and member of the International Program and National Organizing Committees. There was a strong co-operation between CEAUL and INE, the {\it National Statistical Institute},  and {\it Banco de Portugal}, with a high recognition of the field of {\it Extremes} (see \cite{G+P+S-2007c}, and \cite{G+PM+S-2008h}). Such a recognition led to an invited volume of {\it Revstat---Statistical Journal} (Volume {\bf 6}:1, 2008), edited by Jan Beirlant, Isabel Fraga Alves and Ross Leadbetter (see \cite{B+FA+L-2008b}).
The ISI WSC 2007, whose organization was undoubtedly at the genesis of the honour granted to me by the ISI, during the 59th ISI WSC, in Hong-Kong, August 25-30, 2013, with the   `{\it {\rm ISI} Service Award for outstanding contributions to the {\rm ISI} as Chair of the Local Program Committee for the $56$th International Statistical Institute World Statistical Congress {\rm (ISI WSC)}, held in Lisbon, Portugal'}. And the organization of the ISI WSC in 2007 was possibly also  at the genesis of my election as Vice-President of the ISI in December 2014, for the period from July 2015 to July 2019, in my election as corresponding member of the ``{\it Academia das Ci\^{e}ncias de Lisboa}’’ (ACL), in May 2015, and in the invitation to the Scientific Committee of ISI WSC 2017, which took place in Marrakech, July 16--21, and where I organized a session promoting partnerships between the different `{\it National Statistical Societies}' from the ISI (see \cite{MIG-2017b}).

\item 
{\it Workshop on Risk and Extreme Values in Insurance and Finance}, Lisboa,   June 6-7, 2011. 
This conference represented a unique event which brought together in Portugal the three authors of the book {\it Modelling Extremal Events for Insurance and Finance} (see \cite{E+K+M-1997}) --- Paul Embrechts (Zurich, Switzerland), Claudia Kl\"{u}ppelberg (M\"{u}nchen, Germany), and Thomas Mikosch (Copenhagen, Denmark).
This international workshop ran under the sponsorship of CEAUL and the project `EXTREMA: {\it Extremes in Today's World}', PTDC/MAT/101736/2008 (2010-2013), under the leadership of M. Ivette Gomes, and the project `{\it Extremes in Space}', PTDC/MAT/112770/2009 (2011--2013), under the leadership of Laurens de Haan (see \cite{FA+G+H+N-2011}).

\item 5th {\it International Conference on Risk Assessment} (ICRA5), May 29--31, June 1, Tomar, 2013. After 4 meetings  organized by the {\it  International Statistical Institute}--{\it Committee on Risk Analysis} (ISI-CRA),   in Athens, Santorini, Porto Heli and Limassol, 
ICRA5 was organized as part of the {\it Celebrations of the International Year of Statistics} and was held   at the ``{\it Instituto Polit\'{e}cnico de Tomar}’’ (IPT). This meeting was organized  in honour of  Lutz Edler, under the umbrella of CEAUL, UAb---Universidade Aberta, IPT and EXTREMA FCT project  (see \cite{O+G+K+O+G-2013}), and  gave rise to an invited volume of {\it Revstat---Statistical Journal} (Volume {\bf 14}:2, 2016), edited by Christos Kitsos, Teresa Oliveira and Milan Stehl\'{i}k  (see \cite{K+O+S-2016}).
I here highlight  the relevant role of Teresa Oliveira, currently ‘Chair’ of ISI-CRA, and who I also consider a member of the PORTSEA.

\item 
EVT---{\it Extremes in Vimeiro Today}, Vimeiro, September 8--11, 2013. This international conference was organized by  my colleagues and great friends, Ant\'{o}nia Amaral Turkman, Isabel Fraga Alves and Manuela Neves (see \cite{FA+N-2013-Ed}),  on the occasion of my 65th birthday and  for the celebration of the 30 years of the NATO ASI on SEA, in 1983. This was indeed another of the great PORTSEA milestones.
\begin{itemize}
\vspace{-.5pc}
\item  [---] Of the {\bf 40} participants in SEA 1983, only around  30\% were young (less than 35 years old), and only ten (25\%) were women, seven of whom became (or were) PhD students in Portugal. 
\vspace{-.5pc}
\item   [---] And of the {\bf 81}  participants in EVT 2013, more than 50\% were women, and  more than 40\% were young people, something that I saw as very promising for the future of the area  \dots  
\end{itemize}

 \item 
The {\it Workshop on New Frontiers in Statistics of Extremes} (WNFSE 2020) was organized by the `{\it extremists}' Patr\'{i}cia de Zea Bermudez and Miguel de Carvalho, under the sponsorship  of CEAUL and the Research Project,
 {\it Data Fusion and Calibration Methods for Spatial Risk Analysis} (PTDC/MAT-STA-28649/2017), from  FCT. 
The  WNFSE 2020  was held by the end of  February 2020, in Lisbon, just before we went into confinement, due to COVID-19, and left me extremely satisfied and grateful, seeing that the PORTSEA is still very much alive, and with the very active collaboration of several researchers from {\it Banco de Portugal}  who can have a strong effect in the development of the PORTSEA.

 \item A similar comment applies to the organization of the II {\it Institute of Mathematical Statistics} (IMS) {\it International Conference on Statistics and Data Science} (ICSDS) 2023, with 475 presentations, and a high  number of participants. The IMS ICSDS 2023 was held in Lisbon, Portugal, 18-21 December. I was  Co-Chair of the {\it Local Arrangements Committee} (jointly with Eunice Carrasquinha and Teresa Oliveira, both from CEAUL) and member of the {\it International Program Committee}. There was a strong co-operation between IMS and CEAUL,   with a high recognition of the field of {\it Extremes} (see \cite{G+O+O+P+X-2023}). 
 \end{enumerate}

\noindent
Internationally, I refer only  the following:
\begin{itemize}
\vspace{-.5pc}
\item   [---] All EVA conferences, organized since 1998, have had members of our PORTSEA in their Scientific Committees. 
\vspace{-.5pc}
\item   [---] KLIMATEXT---{\it International Conference on Precipitation Extremes in a Changing Climate}, Technical University of Liberec, Hejnice, Czech Republic, September 24-26, 2013. This conference ran under the umbrella of CZ.1.07./2.3.00/20.0086, `{\it Strengthening International Cooperation of the Klimatext Research Team}' (EU Project:  EU structural funds through the Czeck ministry of Education, 2012--2014), with Jan Picek as co-ordinator, and Isabel Fraga Alves and I were two of the  international experts. 
\vspace{-.5pc}
\item   [---] The ``{\it Centre International de Rencontres  Math\'{e}matiques}’’ (CIRM) {\it International Conference on Extreme Value Theory and Laws of Rare Events}, which took place in  July 14-18, 2014, Marseille, France,  had two other prominent members of the PORTSEA in the Organizing Committee, both from the University of Porto, Ana Cristina Moreira Freitas and Jorge Milhazes Freitas, who is a Corresponding Member of ACL since January 2020, and considered by many, and also by me, one of the founders   of {\it Extremes in Dynamical Systems}. 

\vspace{-.5pc}
\item   [---] The 7th {\it International Conference on Risk Analysis} (ICRA7), which took place in Chicago, in May 2017, had Teresa Oliveira in the Executive Committee. ICRA7 was held in my honour, and as {\it Risk Analysis} is not my main research topic, this international tribute had a special flavour, as it seemed more like a global recognition of my contribution to {\it Probability and Statistics}.
\end{itemize}

\subsection{The `heart' of PORTSEA\label{Sec-2.6-Heart}}

\begin{figure}[ht]
 \begin{center}
 \includegraphics[scale=.45, bb=0 0 579 445]{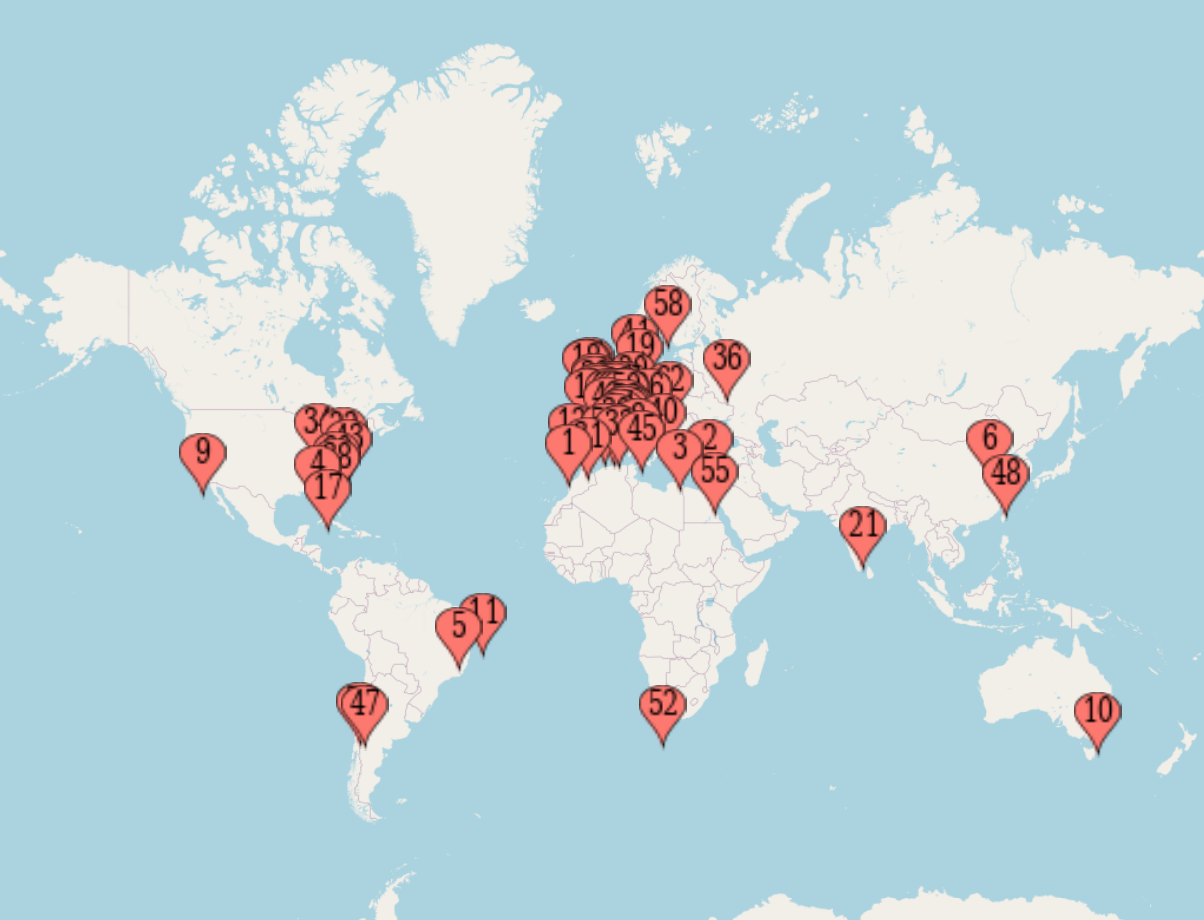}
 \end{center}
 \vspace{-.5pc}
 \caption{\small Universities around the world with co-authors of members of the PORTSEA}
 \label{Internationalization-English}
 \end{figure}

I consider that the excellence of the PhD students that we have had is what has contributed the most to the internationalization of the PORTSEA. But I cannot fail to mention the importance of the co-orientation of a great diversity of foreign graduate students, from different Universities: Charles University Prague, Fudan University of Shanghai, KULeuven,  Pierre-et-Marie-Curie, Siegen, among others.
Over the years, and the increasing number  of publications, collaboration with researchers from other countries has increased significantly, as can be seen in Figure \ref{Internationalization-English}, already provided in \cite{MIG-2021, MIG-2022b, MIG-2023a, MIG-2023b}. We there  represent {\bf 65} universities to which belong co-authors of PORTSEA members, in articles published in high-profile scientific journals by the end of 2020. 

\vspace{.25pc}
The involvement of various members of the PORTSEA as Editors  of scientific journals is another source of satisfaction. Among the wide range of  international periodicals where members of the PORTSEA have played an important role, I would like to highlight only:
\begin{itemize}

\item   [---] The fact that, by the end of 2023, four of these members, Ana Ferreira, Laurens de Haan, Miguel de Carvalho  and the author, were in the body of Associate Editors of {\it Extremes}, the most prestigious journal in the area, edited by Springer, with Thomas Mikosch as current Editor-in-Chief (since 2015), following Holger Rootz\'{e}n  (1998--2006) and J\"urg H\"usler (2007--2014). 

\item   [---] And the fact that I was Editor-in-Chief of   {\it Revstat--Statistical Journal}, from 2003 until  the end of 2018 (a position  held up to the end of 2023 by another member of the PORTSEA, Isabel Fraga Alves), having managed to place this periodical, edited by  INE, with whom we have had high collaboration, among the prestigious journals of {\it Statistics}, with an impact factor in the  ISI {\it Web of Knowledge} since 2007. And the following Editor-in-Chief of {\it Revstat}, for the period 2024--2028, is   Manuel Scotto, another member of the PORTSEA.
\end{itemize}

Again thinking only on the aforementioned periodicals, {\it Extremes} and {\it Revstat}, I  would like to mention the following numbers, reproduced from \cite{MIG-2023a}: Up to the end of 2022, I could find {\bf 46} articles  published at {\it Extremes}, the edition of {\bf 3} issues of {\it Revstat}, {\bf 39} articles published at {\it Revstat} and  {\bf 5} editorial notes, co-authored by members of the PORTSEA.
\vspace{.25pc}

The PORTSEA has nowadays several internationally renowned names. I counted more than {\bf 50}  PhD theses in the area or in very close  areas, written by Portuguese researchers, and associated with degrees obtained in Portugal and abroad. 
But, just as I mentioned before, the number of Habilitation Degrees  needs to increase. 
The current number of PhD and Master students in the area, although not as high as a decade ago, still promises to expand the group soon. 
However, such a number  is slightly decreasing, being thus more difficult to widen the group in the near future, unless our policy is slightly changed. 
 But the dynamic of publication has been quite high, clearly above  international average standards, with more than five hundred articles published in prestigious international journals, as can be seen in \cite{MIG-2023a}. 
\vspace{.25pc}

It should also be noted that our {\it School of Extremes}, despite the high contribution at an international level, has not neglected publication at a national level. This contribution can be attested to by the publication of articles in Portuguese, in the different text collections associated with SPE Congresses and edited by SPE and INE since 1992, where the production in the area of {\it Extremes} has been, on average, 17\% per volume,  in ``{\it Mem\'{o}rias da {\rm ACL}}’’   and in the SPE and SPM Bulletins. For details, see again \cite{MIG-2023a}.
\vspace{.25pc}
     
And  I cannot also fail to mention five relevant books, with PORTSEA members among the co-authors:

\begin{itemize} 
\item []
 \noindent\begin{minipage}{0.2\textwidth} 
 \begin{itemize}
 \item []
\includegraphics[width=\linewidth, bb=0 0 545 825]{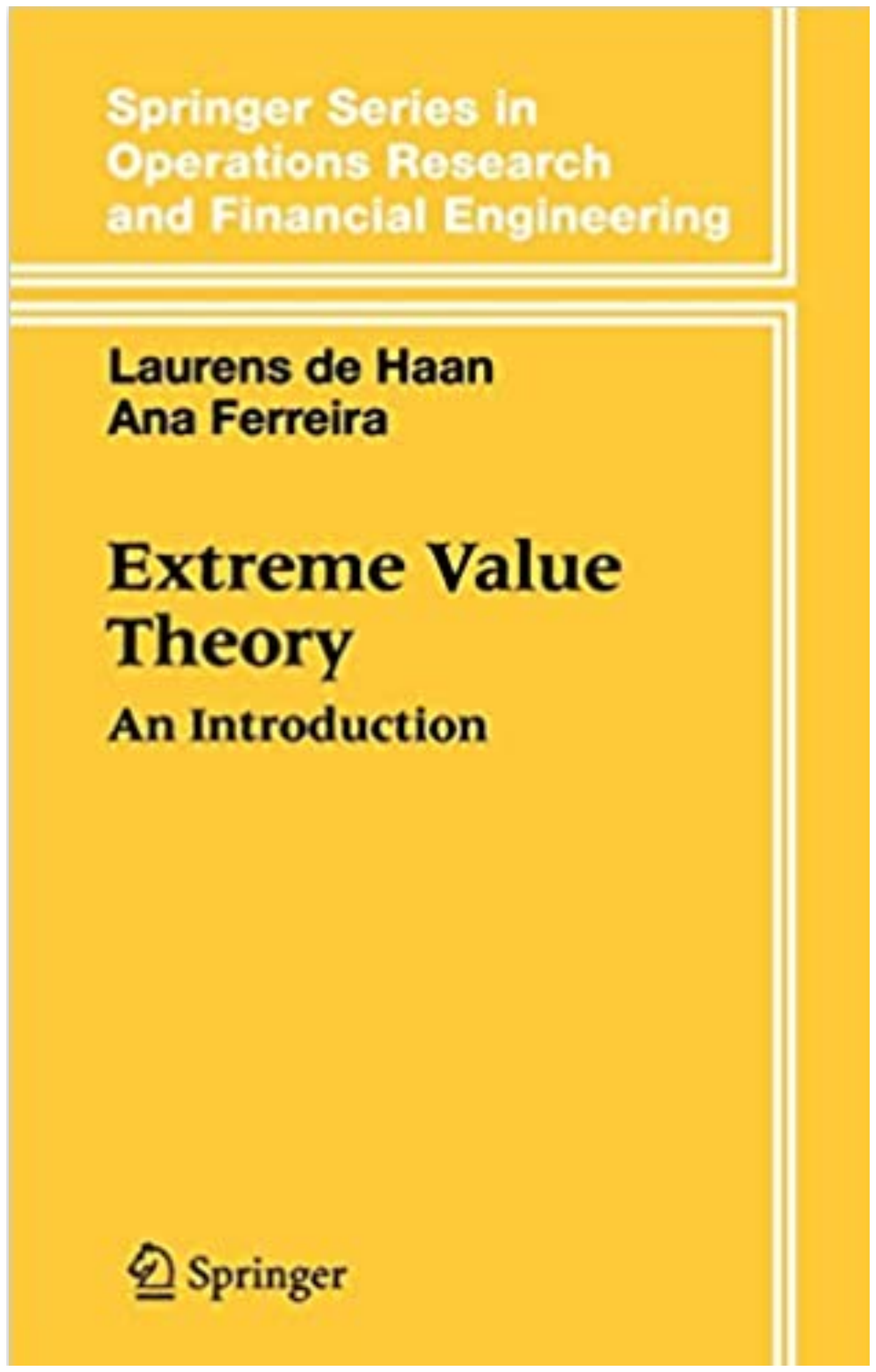}
\end{itemize}
\end{minipage}%
\hfill%
\begin{minipage}{0.7\textwidth}
\begin{itemize}
\item [] A generic book by Laurens de Haan and Ana Ferreira \cite{Haan+F-2006},   with more than 2500 citations, and published by Springer, which once again attest to the invaluable contribution of Laurens de Haan to the establishment of the PORTSEA;
\end{itemize}
\end{minipage}
\vspace{.5pc}

 \noindent\begin{minipage}{0.2\textwidth} 
 \begin{itemize}
 \item []
\includegraphics[width=\linewidth, bb=0 0 577 825]{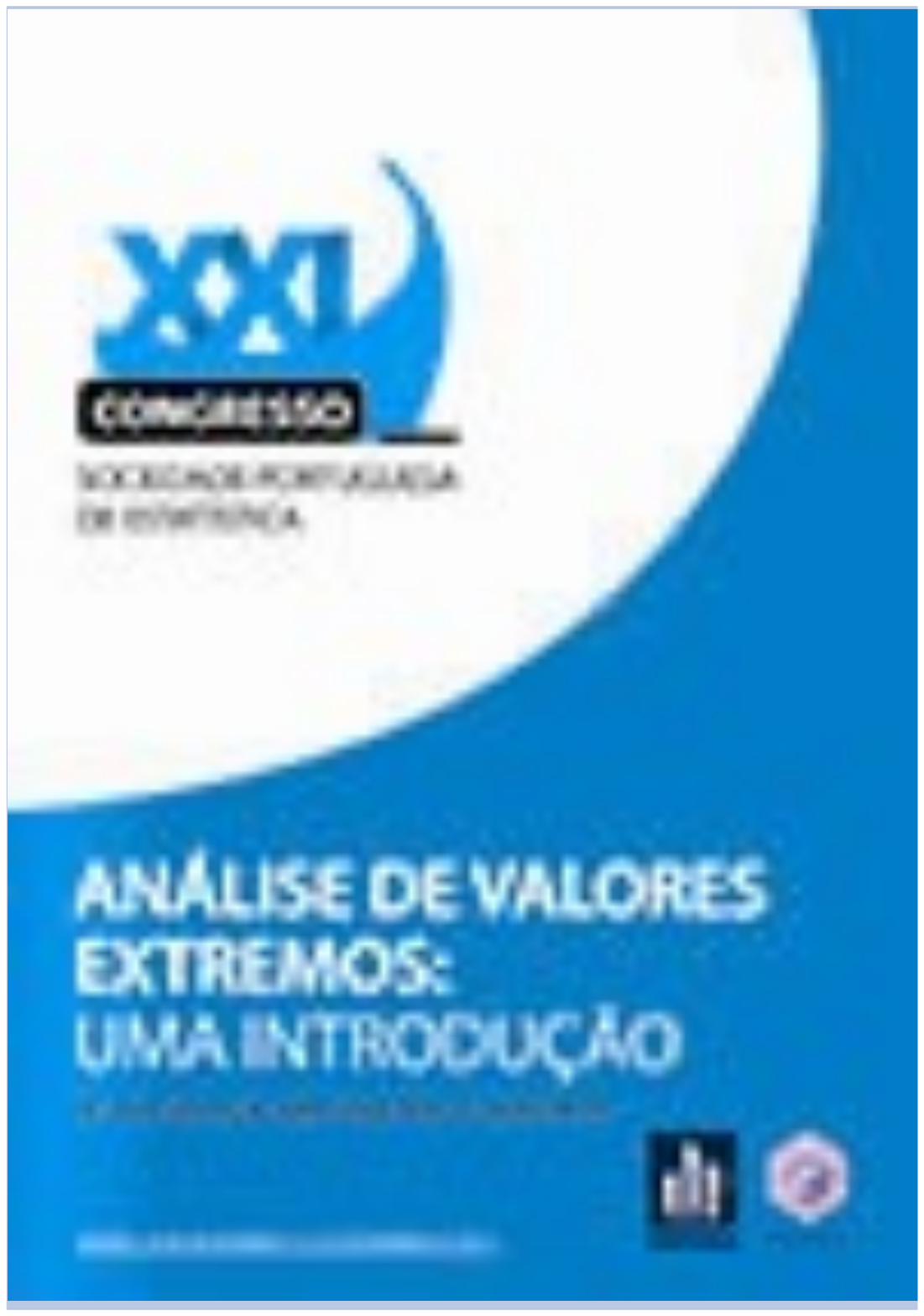}
\end{itemize}
\end{minipage}%
\hfill%
\begin{minipage}{0.7\textwidth}
\begin{itemize}
\item [] A book in Portuguese, co-authored by Ivette Gomes, Isa\-bel Fraga Alves and Cl\'{a}udia Neves \cite{G+FA+N-2013d}, also generic, edited by SPE/INE and associated with a short course taught prior to the XXI {\it Annual Congress of} SPE, held in Aveiro;
\end{itemize}
\end{minipage}
 \vspace{.5pc}

 \noindent\begin{minipage}{0.2\textwidth} 
 \begin{itemize}
 \item []
\includegraphics[width=\linewidth, bb=0 0 503 802]{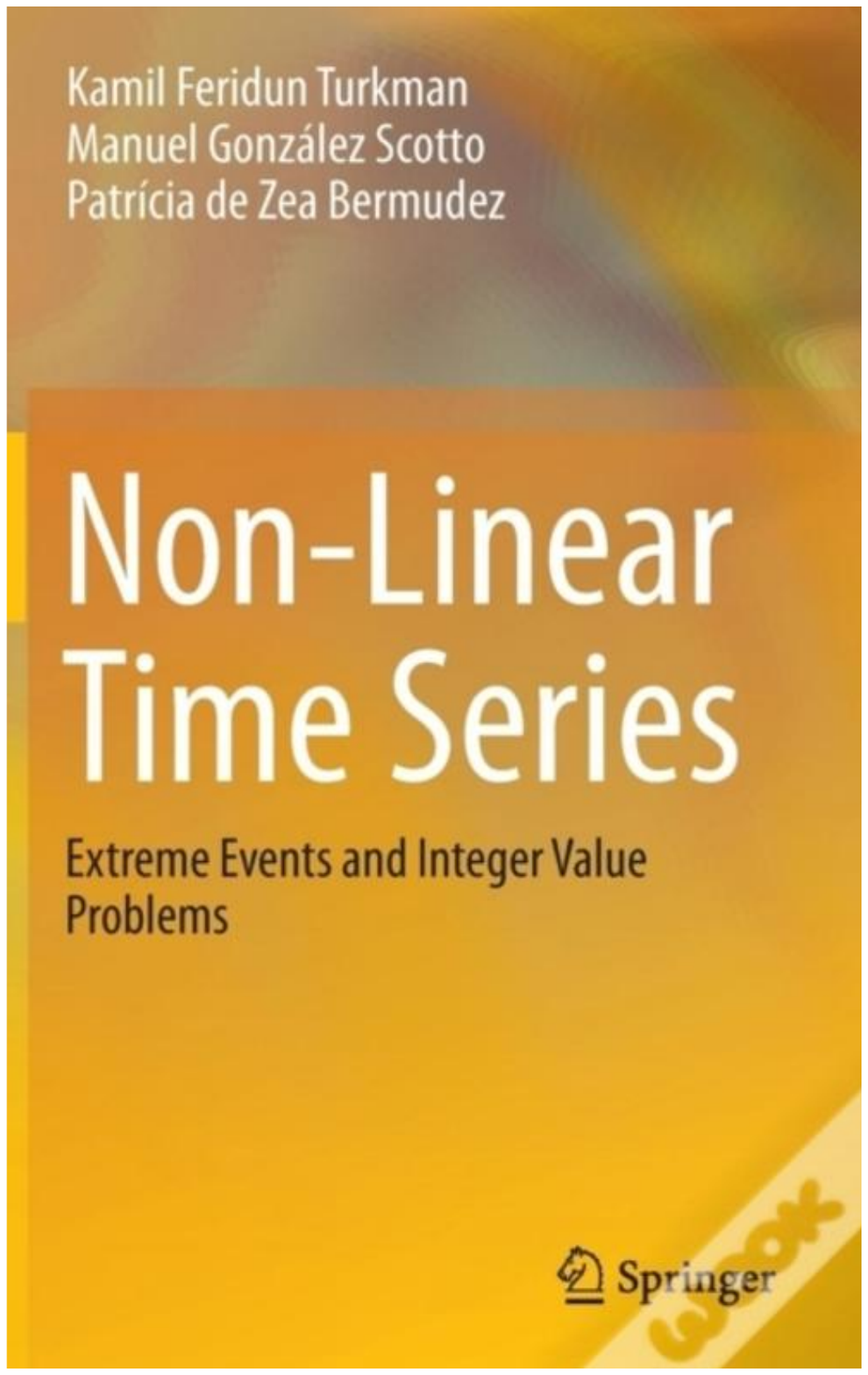}
\end{itemize}
\end{minipage}%
\hfill%
\begin{minipage}{0.7\textwidth}
\begin{itemize}
\item [] A book on \textit{Extremes of Nonlinear Time Series}, co-au\-thored by Feridun Turkman, Manuel Scotto and Pa\-tr\'{i}\-cia de Zea Bermudez \cite{T+S+ZB-2014}, edited by Springer;
\end{itemize}
\end{minipage}

 \noindent\begin{minipage}{0.2\textwidth} 
 \begin{itemize}
 \item []
\includegraphics[width=\linewidth, bb=0 0 547 825]{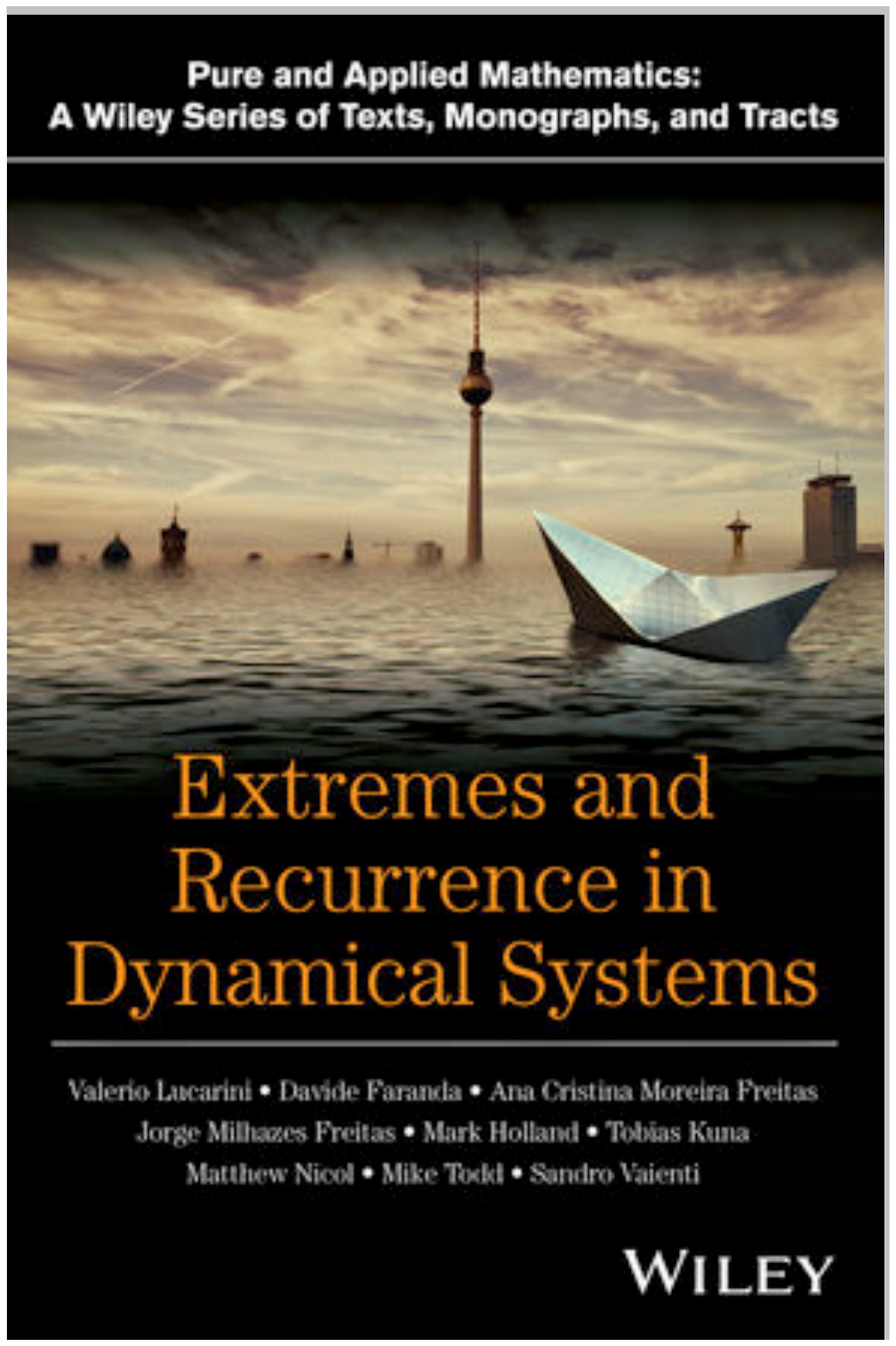}
\end{itemize}
\end{minipage}%
\hfill%
\begin{minipage}{0.7\textwidth}
\begin{itemize}
\item [] Another one about {\it Extremes in Dynamic Systems} \cite{L+F+F+F+K+H+N+T+V-2016}, edited by Wiley, in which Ana Cristina Moreira Freitas and Jorge Milhazes Freitas are co-authors; \end{itemize}
\end{minipage}
 \end{itemize}

\begin{itemize}

\item []

 \noindent\begin{minipage}{0.2\textwidth} 
 \begin{itemize}
 \item []
\includegraphics[width=\linewidth, bb=0 0 545 825]{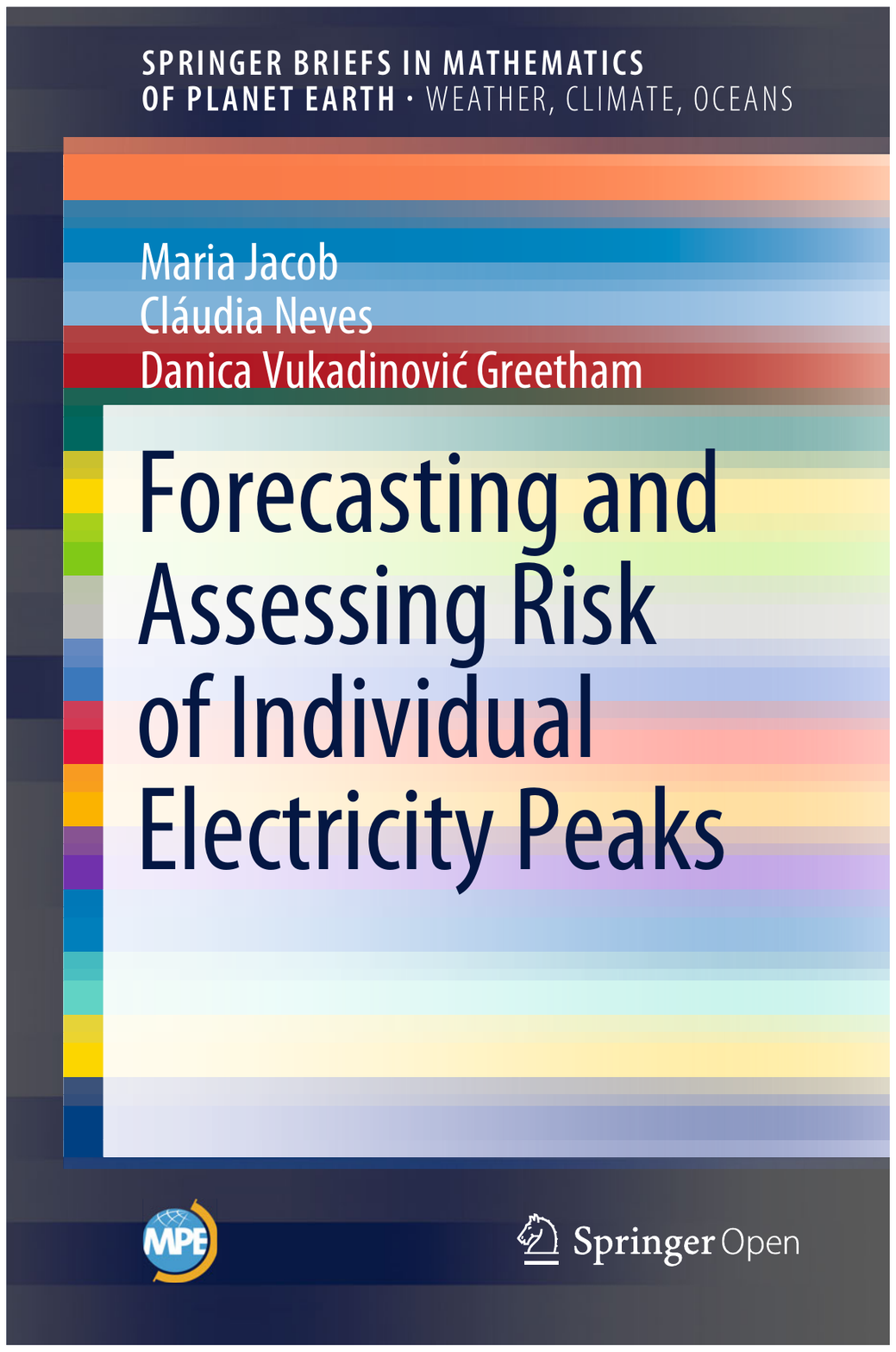}
\end{itemize}
\end{minipage}%
\hfill%
\begin{minipage}{0.7\textwidth}
\begin{itemize}
\item [] And a last one about {\it Risk Assessment and Extremes}  \cite{J+N+G-2020}, edited by Springer, in which Cl\'{a}udia Neves is co-author.
\end{itemize}
\end{minipage}
 
\end{itemize}
\vspace{2pc}

\section{A few details on my scientific research\label{Sec-Personal}}

\subsection{Brief notes about the beginning of my  career\label{Sec-3.1-Beginning}}
In view of what I wrote in \cite{MIG-2014a}, on the occasion of the centenary of Professor Sebasti\~{a}o e Silva, in 2014,  I  shall first vaguely remember, in {\bf \ref{Student-FCUL}},  the times when I was a student at FCUL, between 1965 and 1970, now when I stopped being a teacher more than ten years ago.  I shall next refer, essentially in {\bf \ref{Teacher-FCUL}},  the motivation I felt to move forward with research in the area of  {\it Probability and Statistics}.   I shall next  address, in {\bf \ref{Sheffield}},  the preparation phase of my PhD in Sheffield, mentioning, in {\bf \ref{DeVolta}}, our return to Portugal at the end of 1978, and the two important events in the {\it  History of Statistics} in Portugal. 
Finally, in {\bf \ref{Additional}}, the topics to be approached from {\bf Section \ref{Sec-Inv-Prob}} onwards are put forward.

\subsubsection{A student at FCUL\label{Student-FCUL}}
These times give me a nostalgic longing of a golden age, when everything seemed to be fine, despite the terrible dictatorial political regime in which we lived and against which we fighted.
We expected a lot from our professors, who rarely let me down, and I believe that  professors also expected a lot from us, because at that time `{\it we were going to the university motivated by a great interest in learning}’.
If the memory does not betray me, each professor had her/his own discipline, although  in general she/he had a second discipline, an additional job, which  was paid, although very little.
I may be wrong, but I believe that the bureaucratic tasks at that time did not come close to the levels that currently ruin the lives of university professors and which led me to bring forward an early request for retirement.
\vspace{.25pc}

After the initial six semesters that gave us a solid basic training, the courses took off, and in the last four semesters it was more clear the involvement, I would say even  enchantment, that the professors had for the subjects they were teaching.
Despite having widely enjoyed the course of  {\it Introduction to  Probability and Statistics}, taught by Professor Tiago de Oliveira, and where I realized that `{\it not only the dream but also the chance are constants in life}’, 
I embarked on the option,  {\it Pure Mathematics} (versus  {\it Applied Mathematics}, chosen by the vast majority of my colleagues and friends).
\vspace{.25pc}

The rigour of the courses I had with Professors Almeida e Costa, Dias Agudo,  Lu\'{i}sa Galv\~{a}o, among many others, prepared me for  the courses of  Professor \mbox{Sebasti\~{a}o} e Silva and his direct disciple, Professor Santos Guerreiro, with whom it was very visible the building of  {\it Mathematics}.
It was in these  courses  that I was dazzled by the spirit of {\it Mathematics}, because I then realized better that in {\it Mathematics} creativity and imagination coexist with the rigour of complete and elegant theories.
Not only were these advanced courses better suited to presenting these facets, as also  the preparatory work supported by my previous professors had given me the opportunity to appreciate the role of the hypothesis in {\it Mathematics}, and therefore the role of the guided intuition that helps   mathematicians to explore  new horizons.
\vspace{.25pc}

In one of the two {\it Higher Analysis} courses that I had with Professor Sebasti\~{a}o e Silva, an approach to the  {\it  Theory of Distributions} was natural.
The exposition was fascinating, and without any effort the abstraction of {\it Mathematics} was  undoubtedly associated with the  applications.
That sealed my fate, because until  then I had been   drawn to {\it Algebra}, and I suddenly became fascinated with the applicability of {\it Mathematics}  to the understanding of concrete phenomena. And the choice of optional courses in {\it Stochastic Processes}, taught by Professor Tiago de Oliveira, and in  {\it Probability}, taught by Professor Pedro Brauman, together with the reading of Gnedenko's `super-cheap' Mir’s book, \cite{Gnedenko-1969},   led me to decide to take the option of {\it Applied Mathematics}  in the near future.
Indeed,  in the last two years of my degree in {\it Pure Mathematics}, specialty of {\it Algebra}, I felt fascinated by the little that I had learned about {\it Statistics},  considered here in a broad sense and encompassing {\it Probability} and   {\it Stochastic Processes}.
Despite an attractive invitation from Professor Almeida Costa, an eminent Algebraist at FCUL, to go directly to the {\it United States of America} (USA), with a scholarship from {\it Calouste Gulbenkian Foundation}, I made the decision of going to teach at a family member's college and continuing to study, in order to obtain a degree in {\it Applied Mathematics}, at FCUL, and if possible, to later embark on research into topics in the area of {\it Probability \& Statistics}.
My decision  reached, `through doors and transoms’, the ears of Professor Tiago de Oliveira, a Professor at the {\it Department of Mathematics} at FCUL and a  {\it Statistician} of international renown, who immediately contacted me, convincing me to join as an {\it Assistant}  of the {\it Applied Mathematics Section} at FCUL, where I could study the topics that I liked in a category of `{\it student/professor}’.

\subsubsection{Beginning of teaching and research, also at FCUL\label{Teacher-FCUL}}

I have no doubt that the exceptional merit of some of the Professors I had at FCUL is in the genesis of the professional success that I achieved. 
Among many others, who I will not mention here, I only refer the three professors who clearly drew me towards the university teaching and research: Professor Sebasti\~{a}o e Silva, Professor Almeida e Costa and Professor Tiago de Oliveira.
Professor Sebasti\~{a}o e Silva, who I consider to have been one of the most influential {\it Portuguese Mathematicians} of the 20th century, was  greatly responsible  for the development of my mental autonomy and my critical sense.
And I still remember today a lesson that he gave us and that I often follow: `{\it More important than reading a Mathematics book `from thread to wick’, is to leaf through the book, highlighting only the essential steps and trying to reconstruct many of the omitted ones}'.
According to him, that was the way he had read Laurent Schwartz's book on the {\it Theory of Distributions}, from which he had collected the decisive theorems. 
And it was perhaps the way he was captivated by the subject that allowed him to advance with an axiomatic of such a theory, which, as far as I know,  was the first recorded as complete in the {\it History of Mathematics}, and which has been the starting point for relevant developments by the most diverse researchers.
Professor Almeida e Costa, who considered me `{\it one of the girls who  looked like a man}', encouraged me to briefly study {\it Algebra} in more depth, and he was a Professor that  I certainly disappointed, because, as I mentioned before, I refused his invitation to go to USA, with a PhD scholarship from the  {\it Calouste Gulbenkian Foundation}, and I accepted another invitation from Professor Tiago de Oliveira  to join the  FCUL Section  of {\it Applied Mathematics}, in that happy phase of expansion of universities that Veiga Sim\~{a}o, Minister of Education   between 1970 and 1974, cherished, and where I immediately had to prepare courses on {\it Stochastic Processes}, {\it Simulation} (without never having previously used a computer \dots) and {\it Probability Theory}.
\vspace{.25pc}

It was in July 1970 that I got my degree in {\it Pure Mathematics}, with a specialty in {\it  Algebra}, at FCUL.
 As   already mentioned in {\bf Section \ref{Sec-2.1-Prior}}, I decided to embark on the area of  {\it Probability, Statistics and Stochastic Processes}, 
 and started  working on  {\it Randomness Tests and Non-parametric Methodologies}, under the guidance of Professor J. Tiago de Oliveira, already  at the time an international reference in the  area of {\it Statistics of Extremes},   but which had also started in {\it Algebra}, like me, obviously by mere chance.
During the period that I stayed in Portugal as an Assistant (1970--1975), I began my research first at  CMA, and  next at CEAUL. 
  During this period I managed to co-author two articles (see \cite{A+B+C+G+M+VO-1974} and \cite{G+B+A-1975}), outside the area of {\it Extremes}.
  But despite not having worked directly in the area of {\it Extremes}, I immediately realized some of the potential and beauty of EVT.

\subsubsection{PhD work in Sheffield\label{Sheffield}}

As I mentioned previously, 
Dinis Pestana, my husband, and I obtained doctoral scholarships from the {\it Calouste Gulbenkian Foundation} and were accepted at the {\it University of Sheffield}, UK.
Upon advice of Tiago de Oliveira, I was supposed to work in the area  of {\it Goodness-of-Fit Tests and Non-Parametric Statistics}, under the supervision of Joe Gani.
But, Joe Gani, who I only had the pleasure of meeting in 2005, during my participation in the {\it $55$th Session of the International Statistical Institute}, in Sydney, had  left for Australia, his homeland, just a few days before our arrival in Sheffield.
Provided the chance to be a professor in Sheffield, Clive Anderson, who got to know well Tiago de Oliveira, and his relevant role in the area of  EVT, since he had defended his PhD thesis in 1971, in that  area (see \cite{Anderson-1971}), at the {\it Imperial College of London}.
Clive Anderson  became my PhD supervisor, leading me to investigate on upper  OSs, records, speeds of convergence to limiting laws, pre-asymptotic behaviour, concomitants of OSs --- in short, the `{\it circus}’ of what it was then the research on {\it Extremes} --- themes that served as the basis for writing my PhD thesis.
That's how I ended up embarking, almost `{\it full-time}', in the  area of {\it Extremes}.
I finished my PhD thesis in 1978 (see \cite{MIG-1978}).
\vspace{.25pc}

\subsubsection{Back to Portugal in 1978\label{DeVolta}}
 Dinis Pestana and I returned to Lisbon, already with our PhD degrees, at the end of 1978, still for the {\it Applied Mathematics} Section of the {\it Department of Mathematics} at FCUL.
 And at the beginning of 1979, I was commissioned by Professor Tiago de Oliveira to launch an emerging area in the Lisbon group, {\it Computational Statistics},  an area currently well established in DEIO, FCUL.
To reinforce the degree in {\it Applied Mathematics}, I was also regent of  {\it Inferential Statistics}, {\it Probability} and {\it Stochastic Processes}.

In the 1980/1981 academic year, Feridun Turkman  joined us, at FCUL and also at CEAUL.
As already mentioned, in {\bf Section \ref{Sec-2.1-Prior}},  we therefore had the opportunity to witness two of the important milestones in the development of {\it Statistics} in Portugal. One of these milestones was undoubtedly the founding of the  SPEIO, in 1980. The other milestone was the creation of the DEIOC, in 1981. I was fortunate to be an active element  in the constitution of both DEIOC and   SPEIO, although I cannot fail to mention the primordial and pioneering role of J. Tiago de Oliveira. And upon our proposal, the NATO ASI on SEA took place in the summer of 1983.
At the time, research in mathematics was usually more solitary than it is today. The fashion for international collaborations was not yet in place. Anyway, the publication of research  together with Martin van Montfort, from Wageningen University, in \cite{M+G-1985} and \cite{G+vM-1986}, and my collaboration with Laurens de Haan, from the Erasmus University of Rotterdam,  were certainly auspicious consequences of this long congress. Regarding  joint publications with Laurens de Haan, I am merely referring   to articles in international periodicals  
(see \cite{G+H-1999, FA+G+H+L-2001, G+H+P-2002a, FA+G+H-2003, G+H+P-2004b, G+H+P-2006a, FA+G+H+N-2007a, FA+G+H+N-2009, G+H+HR-2008c}) and to the editions of  books associated with  international conferences organized by us (see \cite{G+H+P+CC+FA-2003, FA+G+H+T-2007b, FA+G+H+N-2011}).
\vspace{.25pc}

\subsubsection{Some additional comments\label{Additional}}
I will next elaborate on what was said in \cite{MIG-2007a}, regarding some of the fields of EVT, in which PORTSEA's contribution has been relevant, with work recognized and published internationally.
These fields are very diverse (see \cite{MIG-2023a}, for further details). 
In addition to a vast group with innovative work in the area of {\it Parametric, Semi-Parametric and Non-Parametric Estimation} of several {\it Parameters of Extreme Events}, univariate and multivariate, PORTSEA has strong groups in different areas of  {\it Extremes and Risk Assessment}.
Giving emphasis to the topics in which I have developed scientific research, 
 I will start with  the development of a few probabilistic results,  in {\bf Section \ref{Sec-Inv-Prob}}.
Always with some underlying probabilistic nature, in  {\bf Section \ref{Sec-Inv-Estat-Uni}},  I  shall move on to {\it Statistics of Univariate Extremes}, under a parametric framework.
In  {\bf Section \ref{Sec-8-4-Semi}}, the  {\it Estimation of Parameters of Rare Events} will be addressed under semi-parametric and non-parametric frameworks, without paying attention to bias-reduction.
The use of resampling methodologies and bias reduction in statistical EVT is the topic to be addressed in {\bf Section \ref{Sec-Jackk+Boot}}. After the presentation of a few generalities in {\bf \ref{Sec-RB-Generalities}}, initial steps in bias reduction, with  emphasis on the jackknife methodology, are provided in {\bf \ref{Sec-Initial-RB+Jackknife}}.
The role of the bootstrap methodology in statistical EVT and its deep use in the estimation of the {\it optimal sample fraction} (OSF) is enhanced in  {\bf \ref{Sec-Bootstrap-only}}.
The semi-parametric estimation of second order parameters is addressed in {\bf  \ref{SubSec-Ordem2}}.
Several additional  methods of  reducing bias are provided in {\bf  \ref{Sec-VIes-Reduzido}}.
 In {\bf Section \ref{SubSec-PORT}}, developments on the PORT methodology are considered, where PORT stands here for `{\it Peaks Over Random Thresholds}’.
Extremes under non-regular frameworks are sketched in {\bf Section \ref{Sec-Non-Regular}}.
Finally, I will give relevance to {\it Applications of} EVT in {\bf Section \ref{Sec-Aplicacoes}}.

\subsection{Main probabilistic results under research\label{Sec-Inv-Prob}}

I shall here distinguish a few topics.

\subsubsection{Probabilistic behaviour of upper OSs\label{Sec-upperOSs}}
I first refer an article, partially out of the scope of my PhD thesis, on the use of fractional calculus in  {\it probability theory} and in EVT (see \cite{G+P-1978}), with  developments in \cite{B+G+P-2011}, where Boop functions and BetaBoop RVs were initially introduced. Their relevance in the investigation of forms of extreme growth has been carried out in \cite{B+G+P-2012b, B+G+P-2014a, B+G+P-2019}. 
In this topic, I still mention \cite{M+P+G-2015}, on the behaviour of upper randomly stopped OSs.
See also, \cite{B+G+M+P+P-2023}, a review paper on the above-mentioned  topics. 
\vspace{.25pc}

Essentially  based on results available in my PhD thesis, \cite{MIG-1978},  I next mention another seminal article, \cite{MIG-1981a}, where the joint distribution of upper OSs  and their concomitants was obtained. Indeed, after outlining the most important results of classical probabilistic EVT and of its extensions for several forms of weak dependence, Gumbel’s approach to statistical inference using maxima was enhanced, and  some of its drawbacks were pointed out. Among them, it was mentioned the wasting of information by only considering the maxima of groups of observations, though generally records of the upper OSs were available.
Let us consider a sequence of IID RVs from a CDF $F(\cdot)$. Using the notation
\[
    M_n^{(1)}
    \equiv M_n
    \equiv X_{n:n}
    =\max(X_1, \dots, X_n),
\]
let us further assume that there exist real sequences  $\{a_n\}_{n\geq 1}$ \mbox{$(a_n>0)$} and $\{b_n\}_{n\geq 1}$  and a non-degenerate CDF $G(\cdot)$ such that 
\begin{equation}
\lim_{n\rightarrow\infty} \mathbb{P}(M_n^{(1)}\leq a_n x +b_n)= \lim_{n\rightarrow\infty}F^n(a_n x+ b_n)= G(x),
\label{Eq-M_n-norm}
\end{equation}
for all $x$ in the set of continuity points of $G(\cdot)$.
Then, as proved in the ETT (see {\bf Section~\ref{Sec-2.1-Prior}}), $G$ is necessarily of the type of the  GEV CDF, in \eqref{Eq-EVI-xi}.
Concerning a more efficient use of the information available, the asymptotic form of the joint distribution of the $i$  largest OSs, $\left(M_n^{(1)}, M_n^{(2)}, \dots, M_n^{(i)}\right)=\left(X_{n:n}, X_{n-1:n}, \dots X_{n-i+1:n}\right),$ $i$ a fixed integer, was  obtained,  and in the proofs, the Poisson properties of upcrossings of a specific level $u_n$, dependent on $n\geq 1$, were used, in a way similar to the one considered by Leadbetter in  \cite{Leadbetter-1974}. 
Provided that \eqref{Eq-M_n-norm} holds, the limiting distribution of $\left(M_n^{(1)}, M_n^{(2)}, \dots, M_n^{(i)}\right)$,  linearly normalized, and where $i$ is a fixed integer, must be of the form,
\begin{multline}
\lim_{n\rightarrow\infty} \mathbb{P}  \left(\bigcap_{j=1}^{j=i} M_n^{(j)}\leq a_n x_j + b_n \right)=  G\left(\mathop{\min}\limits_{1\leq j\leq i} x_j\right)\times \\
 \sum_{r_2=0}^1\sum_{r_3=r_2}^2 \dots \sum_{r_i=r_{i-1}}^{i-1} \prod_{j=1}^{i-1} \frac{1}{(r_{j+1}-r_j)!} \left(\log  \frac{G\left(\mathop{\min}\limits_{1\leq k\leq j} x_k  \right)}{G\left(\mathop{\min}\limits_{1\leq k\leq j+1} x_k \right)}    \right)^{r_{j+1}-r_j}~
 (r_1=0).
 \label{Eq-ivar-CDF}
\end{multline}
The right hand side of \eqref{Eq-ivar-CDF} is to be taken $0$ or $1$ if one of the $x_j$’s is such that $G(x_j)=0$ or if all the $x_j$’s are such that $G(x_j)=1, 1\leq j\leq i$, respectively.
The associated joint  {\it probability density function} (PDF) corresponding to the joint limiting CDF of $(M_n^{(1)}, M_n^{(2)}, \dots, M_n^{(i)})$ suitably normalized, given in \eqref{Eq-ivar-CDF}, is thus given by
\begin{equation}
h(x_1, x_2, \dots, x_i)=\left\{
\begin{array}{lll}
g(x_i) \prod\limits_{j=1}^{i-1} \frac{g(x_j)}{G(x_j)}, & \mbox{if} & x_1>\dots >x_i,\\
0, && \mbox{otherwise,}
\end{array}
\right.
\label{Eq-ivar-PDF}
\end{equation}
where $g(x)=G'(x)$, with $G(x)$  one of the three limiting EV distributions (Gumbel, Fr\'echet and max-Weibull), or, more generally, the GEV CDF in \eqref{Eq-EVI-xi}, being often called an {\it extremal} $i$-{\it dimensional} PDF (see \cite{Weissman-1978}).
The result in \eqref{Eq-ivar-CDF} agrees with the ones of  Lamperti, in \cite{Lamperti-1964} and  Dwass, in \cite{Dwass-1964, Dwass-1966}, on extremal processes.
Distributional properties of particular functions of a vector with such a distribution were derived and several estimation techniques for dealing with multivariate samples of such independent vectors, namely by considering the OSs of their first components and their concomitants, were developed.  An approach to EV practice, similar in spirit to Pickands’, in \cite{Pickands-1975}, and which avoids the arbitrary grouping of data, was further sketched. 
Generalizing Gumbel’s {\it annual maxima} method, it is indeed sensible to consider a multivariate sample
$
(\underline X_1 , \underline X_2 ,\dots,  \underline X_r ),\  \mbox{where} \   \underline X_j  = (X_{1j} ,\dots,X_{i_jj} )$, \mbox{$1 \leq j \leq r$}, 
are {\it extremal multivariate vectors}, with a PDF of the type of the one in \eqref{Eq-ivar-PDF}.
In the univariate case, the notion of OS plays a very important role in statistical methods and is clear and unambiguous.
For multivariate samples no reasonable basis exists for a full ordering of the data, but different generalizations of the concept of order can be made in two or more dimensions (see  Barnett, \cite{Barnett-1976}, for a critical review). That was the main reason why we have decided to consider the ordering of the largest values, $X_{1,j}, 1\leq j \leq r$, and to order the vectors according to the ordering of those largest values, without modifying the random vectors $\underline X_j  = (X_{1j} ,\dots,X_{i_jj})$, $1 \leq j \leq r$, in a way similar to the one used by  Watterson, in  \cite{Watterson-1959} for multivariate normal samples.  
David \& Galambos, in \cite{D+G-1974}, considered the case $i_j=2$, $1\leq j\leq r$, in a bivariate normal situation and used the terminology of concomitants of the OSs, the terminology we have also considered.
I further refer a few articles on the development of inferential methods in the aforementioned multivariate and multidimensional models, to be mentioned in {\bf Section \ref{Sec-Inv-Estat-Uni}}. 
From a probabilistic point of view, this result had also continuity in several publications, among which I enhance \cite{MIG-1984a, MIG-1985a}, on concomitants in multidimensional extreme models.

\subsubsection{Asymptotic and pre-asymptotic behaviour in EVT: rates of convergence\label{Sec-Velocidade}}
In \cite{MIG-1978, MIG-1984c},  the speed of convergence and pre-asymptotic behaviour of the sequence of maxima was deeply discussed.
 Indeed,  parametric inference on the {\it right tail function} (RTF), 
$
 \overline F(x):= 1-F(x),
 \label{RTF}
$
 usually unknown, is  often  done on the basis of the identification of $F^n(a_n x+b_n)$ and of $G_\xi(x)$, in \eqref{Eq-EVI-xi}.
And the rate of convergence  can  validate or not the most usual models in {\it statistics of  extremes}.
In  EVT there exists no analogue of the Berry-Ess\'een theorem that, under broad conditions, gives a rate of convergence of the order of $1/\sqrt{n}$ in the CLT.
The rate of convergence either in the max or the min-scheme depends  strongly on the right or left-tail of $F$, on the choice of the attraction coefficients, and can be rather slow, as first detected by  Fisher and Tippett, in \cite{F+T-1928}. These authors were indeed the first ones to  provide a so-called max-Weibull  penultimate approximation for $\Phi^n(x)$, with  $\Phi$ the normal CDF.
As noted by Fisher and  Tippett, for  the normal CDF, $\Phi$, the convergence of $\Phi^n(a_nx+b_n)$ towards $G_0(x)$, the so-called Gumbel CDF,  is extremely slow.
 They have  then shown numerically  that $\Phi^n(x)$ is `closer'  to a suitable penultimate Weibull ($\xi<0$, in \eqref{Eq-EVI-xi}), rather than the Gumbel CDF $(\xi=0$, in \eqref{Eq-EVI-xi}).
 Such an approximation is the so-called {\it penultimate} or {\it pre-asymptotic  approximation} and several penultimate models have been advanced.
Indeed, an important problem in EVT concerns the rate of convergence of \mbox{$F^n(a_n x+b_n)$} towards 
the GEV CDF,  $G_\xi(x)$, in \eqref{Eq-EVI-xi},
  or, equivalently, the  search for  estimates of the difference 
$$d_n(F, G_\xi, x):=F^n(a_n x+b_n)- G_\xi(x),$$
or the finding of $d_n\rightarrow 0$, as $n\rightarrow\infty$, and $\varphi(x)$ such that 
$$F^n(a_nx+b_n)- G_\xi(x)=d_n\varphi(x)+o(d_n),$$ 
for all $F\in {\cal  D_M}(G_\xi)$, the whole max-domain of attraction of $G_\xi$, in \eqref{Eq-EVI-xi}, $\xi\in\mathbb{R}$.
We then say that the rate of convergence of $F^n(a_nx+b_n)$ towards $G_\xi(x)$ is of the order of $d_n$. 
  In this same framework,  the possible penultimate  behaviour of \mbox{$F^n(a_n x+b_n)$} has been   studied, i.e.\  the possibility of finding $H(x)=H_n(x)$, not necessarily  an MS CDF, satisfying \eqref{Eq-MS-equation},  such that
$$F^n(a_nx+b_n)-H_n(x)=O(r_n), \quad r_n=o(d_n).$$ 
 And it is true that this pre-asymptotic behaviour continued to bear fruit, as can be seen in \cite{MIG-1979b, MIG-1986, MIG-1989b}.
 Among others, we further refer an article by Gomes and Pestana in 1986 (see  \cite{G+P-1987}), where it was observed that the penultimate phenomenon in extremes is not in the least characteristic of the domain of attraction of the Gumbel CDF, together with the notion of standard and non-standard domains of attraction, and an article by Gomes and de Haan, \cite{G+H-1999}, in 1999, where it is derived, for all \mbox{$\xi\in\mathbb{R}$},   exact penultimate approximation rates with respect to the variational distance,  and under adequate   differentiability assumptions. 
\vspace{.25pc}
 
The modern theory of rates of convergence in EVT began with  Anderson, in \cite{Anderson-1971}. 
Dated overviews of the topic can be seen in  \cite{Galambos-1984} and in \cite{MIG-1994b},  the first article I wrote in LaTeX, related to an invited lecture given in Gaithersburg, in 1993, at the ``{\it Conference on Extreme Value Theory and its Applications}’’, organized by Janos Galambos, among others.
Further information on the topic can be seen in the overviews by  Beirlant, Caeiro \& Gomes, in \cite{B+C+G-2012} and  Gomes and Guillou, in \cite{G+G-2015}.
\vspace{.25pc}

Recently, and following the PhD thesis by Paula Reis (see \cite{ReisP-2012}), the role of MS penultimate approximations in reliability has been considered in \cite{G+R+CC+D-2013h, G+R+CC+D-2017, R+CC+D+G-2015, MIG-2020}. 
Indeed, any coherent system can be represented as either a {\it series-parallel} (SP)---a series structure with components connected in parallel  or a {\it parallel-series} (PS) system---a parallel structure with components connected in series (see   \cite{B+P-1975}). Its lifetime can thus be written as the minimum of maxima or the maximum of minima. For large-scale coherent systems it can be sensible to assume that the number of system components goes to infinity. Then, the possible non-degenerate EV laws, either for maxima or for minima, are eligible candidates for the finding of adequate lower and upper bounds for such a reliability.
Dealing with regular and homogeneous PS and SP  systems, the above-mentioned  authors have assessed the gain in accuracy when a penultimate approximation is used instead of the ultimate one.
Developments have followed different directions that can be seen  in \cite{MIG-2023a}.
\vspace{.5pc}

Even more recently, the {\it Weibull tail coefficient} (WTC), strongly related to the pre-asymptotic behaviour of extremes associated with a normal model, has been relevant in the research of members from PORTSEA. 
The WTC is the parameter $\omega$ in an RTF of the type
\begin{equation}
{\overline F(x)=1-F(x)=: e^{-H(x)}, \quad H\in {\cal R}_{1/\omega}, \ \omega\in\mathbb{R}^{+},}
\label{Eq-WTC}
\end{equation}
where ${\cal R}_\alpha$ denotes the set of regularly varying functions with an index of regularly variation equal to $\alpha$, i.e.\ positive measurable functions $H$ such that $\lim\limits_{t\rightarrow\infty} H(tx)/H(t)=x^\alpha$, for all $x>0$.
 Equivalently to (\ref{Eq-WTC}), 
 $$
\label{Weibull_tail}
\lim_{t\to \infty} \frac{\ln (1-F(c\,t))}{\ln (1-F(t))}=c^{1/\omega},
$$
for all $c>0$, and also equivalently, we can say that
$$
{U(e^t)=H^\leftarrow (t)\in {\cal R}_\omega \quad \Longleftrightarrow\quad U(t)=: (\ln t)^\omega L(\ln t),}
\label{Eq-CGHR-14}
$$
with $L\in {\cal R}_0$, a slowly varying function (for details on regular and slow variation, see \cite{B+G+T-1987}.
In  a context of EVT for maxima,  it is possible to prove that we have  a null  EVI,  $\xi=0$,  but usually a very slow rate of convergence. 
We are working with those right-tails, like the Normal RTF, in the {\it max-domain of attraction} (MDA) of Gumbel's law $\Lambda(\cdot)$, which can exhibit a penultimate (or pre-asymptotic)  behaviour.
Such RTFs, despite of double-exponential, look more similar either to
\begin{itemize}
\item [---] Max-Weibull, $\Psi_\alpha(x)=\exp(-(-x)^\alpha)$, $x<0$ ($\xi=-1/\alpha<0$)
\item [---] or to  Fr\'echet, $\Phi_\alpha(x)=\exp(-x^{-\alpha})$, $x>0$ ($\xi=1/\alpha>0$) 
 \end{itemize}
 right tails, according as  $\omega<1$ or $\omega>1$, respectively. 
Regarding the WTC-estimation, see {\bf \ref{Sec-Other}}.

\subsubsection{Extremes of dependent sequences and other multivariate structures\label{Sub-Sec-Dependente}} 
Suppose now that $\{X_n\}_{n\geq 1}$ is a stationary sequence of RVs, that is, $\left(X_{i_1}, X_{i_2}, \dots, X_{i_r} \right)$ and $\left(X_{i_1+k}, X_{i_2+k}, \dots, X_{i_r+k} \right)$  are ID for any choice of $r$, $k$ and $i_j$, $1\leq j \leq r$.
Further suppose that the original CDF of each of the $X_i$'s is $F(x)$ and use the same notation as before, $M_n^{(1)}\equiv X_{n:n}=\max (X_1, X_2, \dots, X_n)$.
The ETT or Fisher-Tippett-Gnedenko’s theorem, already mentioned in {\bf Section \ref{Sec-2.1-Prior}},   has been generalized for stationary sequences with different types of ``decay dependence''.
 Watson, in \cite{Watson-1954}, and later  Newell, in \cite{Newell-1964}, have shown that if the sequence $\{X_n\}_{n\geq 1}$ is $m$-dependent, i.e.\  if $X_i$ and $X_j$ are independent whenever $|i-j|\geq m$, then the functional equation verified by the limit distribution of $M_n^{(1)}$, suitably normalized, is the one we find in the independent case and hence the ETT has an obvious analogue.
 Loynes, in \cite{Loynes-1965}, has considered stationary strong mixing sequences, and the lack of independence was again found not to affect the problem and the only possible non-degenerate limit laws of $M_n^{(1)}$ suitably normalized are going to be the same as in \eqref{Eq-EVI-xi}.
Another approach to the problem was given by  Galambos, in \cite{Galambos-1972}.
Again, he considered a set of RVs of which he says ``almost all'' are ``almost independent'' and with some restrictions on the bivariate structure of the stationary sequence, their maximum is shown to behave as if they were independent.
The strong mixing condition may  be weakened in several different ways, and 
Leadbetter, in \cite{Leadbetter-1974},  considered the  so-called  $D(u_n)$ condition.
Under the validity of $D(u_n)$ for the stationary sequence $\{X_n\}_{n\geq 1}$ and for sequences $\{u_n\}_{n\geq 1}$ of the form $u_n=a_n x+b_n$, for every $x$, where $\{a_n\}_{n\geq 1}$ $(a_n>0)$ and $\{b_n\}_{n\geq 1}$  are sequences of real constants such that \eqref{Eq-M_n-norm} holds, an analogue of the ETT can be proved. 
The {stationary sequence}  $\left\{X_n\right\}_{n\geq 1}$ is said to have an {\it extremal index} (EI),  $\theta$ $(0<\theta\leq 1)$,  if, for all $\tau>0$, we can find a sequence of levels $u_n=u_n(\tau)$ such that, with $\left\{Y_n\right\}_{n\geq 1}$ the associated IID sequence (\emph{i.e}.\  an IID sequence from the same CDF, $F$),
\begin{equation}
\mathbb{P}\left(Y_{n:n}\leq u_n\right)=F^n(u_n)\mathop{\longrightarrow}\limits_{n\rightarrow\infty}\ e^{-\tau} \quad  \ \mbox{and}\ \quad   \mathbb{P}\left( X_{n:n}\leq u_n\right)\mathop{\longrightarrow}\limits_{n \rightarrow\infty}\ e^{-\theta \tau}.
\label{Eq-EI}
\end{equation}

For a stationary sequence $\{X_n\}_{n\geq 1}$, and under a dependence condition denoted by \mbox{$\widetilde D\left( u_n^{(i_1)}, \dots, u_n^{(i_{p+q})} \right)$} for suitable $\{u_n^{(j)}\}_{n\geq 1}$, $1\leq j\leq i$, $i_k\in \{1, 2, \dots, i\}$, \mbox{$1\leq k\leq p+q$}, $i_{k+1}-i_k\geq 0$, which is stronger than $D(u_n)$, but weaker than strong-mixing,  Gomes, in \cite{MIG-1978}, has obtained the same  joint limiting CDF, in \eqref{Eq-ivar-CDF}, for the $i$ largest values $(M_n^{(1)},   \dots, M_n^{(i)})$ of $(X_1, X_2, \dots, X_n)$, as $n\rightarrow\infty$, where $i$ is a fixed positive integer.
And several results formerly obtained in \cite{MIG-1978, MIG-1981a} were extended to the case of weakly dependent  RVs (see \cite{MIG-1979a, MIG-1980a}). Almost simultaneously,  Adler, in \cite{Adler-1978}, considered also stationary sequences satisfying dependence restrictions similar to those of Leadbetter. The dependence assumptions  assumed in \cite{MIG-1978} are implied by the ones he assumed, but on the other hand he obtained not only convergence of the joint finite dimensional distributions but also weak convergence of such a process. The methods of proof are however different and were derived independently.
A detailed reference to EVT under dependent frameworks, and the role of the  EI, $\theta$, defined in \eqref{Eq-EI}, a parameter of extreme events related to clusters of extremes in dependent frameworks, and almost as important as the EVI, was given in the 1983 book by Leadbetter, Lindgren and Rootzén (see \cite{L+L+R-1983}).
\vspace{.25pc}

In this area my contribution was only marginal, but apart from the aforementioned articles,    I still mention  \cite{MIG-1993a, MIG-2001} and \cite{G+H+P-2004b, G+H+P-2006a}, relegating some other articles published internationally to the estimation of the EI.
But I am very happy to have been lucky enough to have had PhD students like Teresa Alpuim (see \cite{Alpuim-1989}), Emilia Athayde,  (see \cite{Athayde-1994}), Helena Ferreira (see \cite{FerreiraH-1994}), Andreia Hall (see \cite{Hall-1998b}) and   Cristina Miranda (see \cite{Miranda-2005}), to whom I managed to awaken interest in this area, which I personally always found fascinating, and who  developed work of great value, such as can be seen in \cite{MIG-2023a}.

\subsection{Univariate extremes: Parametric framework\label{Sec-Inv-Estat-Uni}} 

For  reviews of the topic {\it Univariate Extreme Statistics---Parametric vs Non-Parametric Models}, see \cite{C+G-2013c}, written  in Portuguese, and the two international review articles, \cite{B+C+G-2012} and \cite{G+G-2015}.

\subsubsection{Statistical choice of extremal models (the `old Trilemma’ of Tiago de Oliveira)\label{Sec-Trilemma}}

 Let us again consider $G_\xi(x)=\exp(-(1+ \xi x)^{- 1/ \xi} ), 1+\xi x >0$, $ \xi\in \mathbb{R}$,  the  GEV  or von Mises-Jenkinson  CDF, in \eqref{Eq-EVI-xi}. Due to its simplicity, Gumbel distribution, $\Lambda(x)\equiv G_0(x)$, is a favourite one in statistical  EVT and thus, discrimination between EV models, with $ \xi=0$ playing a central and prominent position is an important statistical problem.
 The first work in this field, in which Tiago de Oliveira was a pioneer, was the development and study of {\it locally most powerful} (LMP) tests for discrimination between EV models (see \cite{TiagoO-1981}).
 This theme once again began its development under  a parametric framework and Tiago de Oliveira was  author of several additional articles, mentioned  in \cite{MIG-2023a}, and  was also co-author of computational approaches to  LMP tests (see \cite{TO+G-1984}), where two test statistics were proposed for the choice of univariate EV models and asymptotic properties of the statistic proposed by Gumbel, 
 \begin{equation} 
 W_n(\underline X)= \frac{X_{n:n}-X_{\lfloor n/2\rfloor+1:n}}{X_{\lfloor n/2\rfloor+1:n}-X_{1:n}},
 \label{Gumbel-statistic}
 \end{equation}
 with $\lfloor x\rfloor$ the integer part of $x$, were obtained for testing   $\xi = 0$ in the GEV CDF.
\vspace{.5pc}

It is also sensible mentioning some articles in which I am the author, \cite{MIG-1982, MIG-1984d, MIG-1985b, MIG-1987, MIG-1989a, MIG-1989c}. I was co-author of a few additional articles on this theme: \cite{M+G-1985};  \cite{G+A-1986},   who compared asymptotically the power of two statistics for solving statistical choice in a multivariate GEV ($\xi$) set-up, the Gumbel statistic, in \eqref{Gumbel-statistic}—which balances the upper and lower range of the sample—and  an analogue of the LMP test statistic;
\cite{G+vM-1986}, where  the  exponential CDF is tested versus the GP CDF;
\cite{A+G-1987}, who worked  under Mejzler’s hypothesis (see \cite{Mejzler-1956}) and provided different statistical choice tests.
A few years later, in 1996, Fraga Alves \& Gomes, in \cite{FA+G-1996}, compared with several other connected perspectives a  test procedure  introduced in \cite{H+W-1992}.
This 1996 article, although still with a strong parametric flavour, is an article in which it is possible to notice the existence of an inflection towards the semi-parametric framework. And Laurens de Haan, who had invited me to the “{\it Conference on Multivariate Extreme Value Estimation with Application to Economics and Finance}”, which took place in Rotterdam, in 1994,  had a great responsibility here.
  The most recent contributions to this topic have been developed in a semi-parametric environment and were one of the topics on which Claudia Neves (see \cite{NevesC-2006}) worked for her PhD, under the supervision of Isabel Fraga Alves and Laurens de Haan.

\subsubsection{Univariate statistics of extremes based upon multivariate and multidimensional models} 
This is the phase in which the  statistical EVT  developed in Portugal was still parametric in nature, closely linked to the GEV and GP models, but essentially based on multivariate, multi-dimensional and other non-classical models, together with some incursions into non-parametric models, perhaps too vast to have proved interesting. 
I refer to the articles  \cite{MIG-1984b, MIG-1985c},   where  comparisons of several approaches to the topic were considered, essentially taking into account the PDF in \eqref{Eq-ivar-PDF}. See also \cite{G+A-1986, G+P-1986a, A+G-1987}. Here we were looking for parametric models that would generalize Gumbel's annual maximum method, often  based in \eqref{Eq-ivar-PDF}, being thus more informative.
I also refer  \cite{MIG-1997}, where the concomitants of OSs were used for the estimation of  the correlation coefficient.
\vspace{.25pc}

\subsubsection{Other non-classic models}
Since the end of the 20th century, the class of {\it max-semi-stable} (MSS) distributions introduced, independently, by  Grinevich, in \cite{Grinevich-1992}, and Pancheva, in \cite{Pancheva-1992}, has been much cherished by some Portuguese researchers. Such a class has the functional form,  
\begin{equation*}
{\rm MSS}_{\xi,\nu}(x)=\left\{ \begin{array}{lll}
{\rm e}^{-\nu\{\ln(1+\xi x)/\xi\}(1+\xi x)^{-1/\xi}},\ 1+\xi x>0, & \mbox{if} & \xi\not=0,\\
{\rm e}^{-\nu(x)\exp(-x)},\ x\in\mathbb{R},  & \mbox{if} & \xi=0,
\end{array}
\right.
\label{Eq-MSS}
\end{equation*}
where $\nu(\cdot)$ is a positive, limited and periodic function, being   ${\rm MS}_{\xi}\equiv G_\xi={\rm MSS}_{\xi,1}$.
Statistical inference for these models has been developed slowly along time,  due to the complexity of the topic.
In fact, in \cite{MIG-2005} it is mentioned that the unique published article on statistical inference in MSS models  was of a national scope, \cite{T+G+CC-2001}.
However Sandra Dias worked hard on estimating the parameters of the class of MSS laws for her PhD (see \cite{DiasS-2007}), supervised by Luisa Canto e Castro, having achieved some publications nationally and internationally (see again  \cite{MIG-2023a}).
\vspace{.25pc}

The {\it Birnbaum-Saunders} (BS) model (see \cite{B+S-1969}) is a life distribution originated from a problem of material fatigue that has been largely studied and applied in recent decades. An RV following the BS distribution can be stochastically represented by another RV used as basis. Then, the BS model can be generalized by switching the distribution of the basis variable using diverse arguments allowing to construct more general classes of models.
The development of research work with Victor Leiva, from {\it Pontificia Universidad Catolica de Valparaiso}, led us to consider Birnbaum-Saunders' EV laws, studied from different perspectives in the following international articles: 
\cite{F+G+L-2012, G+F+L-2013a, G+F+L-2013b, L+F+G+L-2016, L+L+G+F-2019}. 
\vspace{.25pc}

Additionally I cannot fail to refer the need to develop methods for estimating the parameters of pre-asymptotic laws for extremes, of the type of the ones mentioned in a discussion started in \cite{G+R+CC+D-2017}, never completed and not yet published.

\subsection{Non-reduced bias semi-parametric and non-parametric estimation of parameters of rare events\label{Sec-8-4-Semi}}
 In this field, where we do not work with specific parametric models, but in broad domains, such as the entire MDA of the CDF $G_\xi$, in \eqref{Eq-EVI-xi}, for which we use the notation ${ \cal D_M}(G_\xi)$, and where $\xi$ is the so-called EVI, the work of our School has been vast.
 
\subsubsection{EVI estimation}
 We begin by mentioning some of the classic estimators of the parameter $\xi$, in a semi-parametric context. Given a sample  \mbox{$\underline {\bf X}_n:=$} $(X_1, \dots, X_n)$, IID, or even stationary, associated with a CDF $F$, let us recall the notation $X_{i:n}, 1 \leq i\leq n$, for the sequence of ascending OSs.
For $\xi>0$, that is, ${\cal D_M}^+:={\cal D_M}(G_{\xi>0})$, the classic EVI estimators are the Hill estimators (see \cite{Hill-1975}), with the functional form
  \begin{multline}
{\rm H}(k) \equiv {\rm  H}(k; \underline{\bf X}_n ) := \frac{1}{k}\sum_{i=1}^k \ln  {X_{n-i+1:n}}-\ln {X_{n-k:n}}=: M_n^{(1)}(k),\\ \quad  1\leq k<n, 
  \label{Eq-Hill}
  \end{multline}
where
  \begin{equation}
  M_n^{(p)}(k):= \frac{1}{k}\sum_{i=1}^k \left\{\ln  {X_{n-i+1:n}}-\ln {X_{n-k:n}}\right\}^p,\quad p\in\mathbb{R},
  \label{Eq-M_n^p}
  \end{equation}
are the moments of order-$p$ of the log-excesses.
We further refer the {\it moment estimator}, introduced and studied in  \cite{D+E+H-1989}, valid for $\xi\in\mathbb{R}$, and with the functional form
\begin{equation}
{\rm M}(k) \equiv {\rm  M}(k; \underline{\bf X}_n ) := M_n^{(1)} +1-\frac{1}{2}\left\{ 1- \left(M_n^{(1)}\right)^2/M_n^{(2)}\right\}^{-1},
\label{Eq-Momentos-DEH}
\end{equation}
with  $M_n^{(p)}$ defined in \eqref{Eq-M_n^p}.

 Once again with under evaluation, regarding the estimation of the EVI and outside the context of reducing bias, to be considered in {\bf Section \ref{Sec-Jackk+Boot}}, and of most of the articles on censoring and truncation and on the role of {\it generalized means} (GMs) in estimating parameters of extreme events, to be considered later, in this  Section, I mention the following articles, placed in a chronological order:
 \cite{M+G+N-1999, M+G+N-2004}, by Martins, Gomes and M Neves; 
\cite{G+O-2003a, G+O-2003b, G+O-2003c}, by Gomes and O Oliveira;
\cite{G+CC+FA+P-2008a}, by Gomes, Canto e Castro, Fraga Alves and Pestana, with the pioneering contributions of Laurens de Haan in {\it Statistics of Extremes} for IID data and advances in the estimation of the EVI;
\cite{FA+G+H+N-2009}, by Fraga Alves, Gomes, da Haan and C Neves, about  the  {\it mixed moment} (MM) estimator. 
The MM estimator is based not only on the order-$p$ moments of the log-excesses, in \eqref{Eq-M_n^p}, but also on
\begin{equation}
L_n^{(p)}(k) :=\frac{1}{k} \sum_{i=1}^k \left(1-\frac{X_{n-k:n}}{X_{n-i+1:n}}\right)^p, \quad p\geq 1,
\label{Eq-L_n^p}
\end{equation}
being defined by
\begin{equation}
{\rm MM}(k) \equiv {\rm  MM}(k; \underline{\bf X}_n ) := \frac{\widehat \varphi(k)-1}{1+2\min\left(\widehat\varphi (k)-1, 0 \right)}
\label{Eq-MM}
\end{equation}
where
$$
\widehat\varphi(k):= \frac{M_n^{(1)}(k)-L_n^{(1)}(k)}{\left( L_n^{(1)}(k)\right)^2},
$$
with $M_n^{(p)}(k)$ and $L_n^{(p)}(k)$ respectively defined in \eqref{Eq-M_n^p} and in\eqref{Eq-L_n^p};
\cite{C+G-2013a}, by Caeiro and Gomes, on  {\it probability weighted moments} (PWM) estimation;
\cite{G+S-2014}, by Gomes and Stehl\'ik, on different modifications of Hill estimators.
Regarding the estimation of the {\it tail index} (a positive EVI)  in dependent structures, I refer a unique article by Gomes and Miranda, 
\cite{G+M-2009}.

\subsubsection{EI estimation}
 For a sequence of  IID RVs any point process limit for the time normalized exceedances of high levels is a Poisson process. But for stationary dependent sequences of RVs, under adequate general local and asymptotic dependence restrictions (see \cite{L+L+R-1983}), any point process limit for the time normalized exceedances of high levels is a compound Poisson process, i.e.\ there is a clustering of high exceedances, where the underlying Poisson points represent cluster positions, and the multiplicities correspond to cluster sizes. For such a class of stationary sequences, beyond the  EVI, there exists and is well defined the  EI, $\theta, 0<\theta \leq 1$, in \eqref{Eq-EI}, which plays a key role in determining the intensity of cluster positions, because it may be proved that for many interesting situations (not in general!) the EI is the reciprocal of the limiting mean cluster size.
The estimation of the  EI (see \cite{L+N-1989}, among others), or more generally inference in dependent processes, are topics that have also had a relevant contribution from our School.
I refer to the articles, \cite{MIG-1990b, MIG-1993a, MIG-1993b, MIG-1995b, MIG-2015}.
I further mention the following recent articles on a robust estimation of the EI, \cite{G+M+SM-2020c, SM+M+G-2022}, in a joint work with Cristina Miranda and Manuela Souto de Miranda.

\subsubsection{Optimal sample fraction (OSF) estimation}
The choice of the  OSF ($k/n$) in classical procedures associated with  statistical EVT  was  investigated in \cite{O+G+FA-2006}, published posthumously as a tribute to Orlando Oliveira, due to the unexpected and premature death of  Orlando, a  doctoral student of mine, shortly after defending his PhD thesis in 2003 (see \cite{OliveiraO-2003}).
I also refer some articles where methods for choosing the OSF, not necessarily using the bootstrap methodology (a topic to be covered in {\bf Section \ref{Sec-Jackk+Boot}}) are devised:
\cite{C+G-2011c, C+G-2015c, C+G-2022, G+P-2011, N+G+F+PG-2015}.

\subsubsection{Estimation of other parameters of extreme events\label{Sec-Other}}
Regarding the estimation of high quantiles or {\it value-at-risk} (VaR), return periods, probabilities of exceedance of high levels, right endpoint of the model underlying the data, among other parameters of extreme events, there was also a relevant contribution from PORTSEA researchers.
I am co-author of the following articles, which focus essentially on VaR estimation:
\cite{G+HR+V-2006b, G+P+HR+V-2008g, G+B+P-2015b, G+C+F-2015c};
I further mention the articles,
\cite{C+G-2011d, C+G+V-2014},  with general methods of semi-parametric estimation of the right tail  function through the PWM method, with PWM standing for {\it probability weighted moments}.
In what concerns the WTC-estimation, I mention, 
\cite{C+G+HR-2022a, C+HR+G-2022b, HR+C+G-2024}.

\subsubsection{Censoring and truncation}
In Statistics of Extremes the most common assumption on any set of univariate data is to consider that we are in the presence of a complete sample. However, in the analysis of some physical phenomena such as wind speed, earthquake intensity or floods, extreme measurements are sometimes not available because of damage in the instruments. Also, in the analysis of lifetime data or reliability data, observations are usually censored.  
The methodologies in \cite{B+G+D+FV-2007} and in  \cite{E+FV+G-2008}  were applied to a few sets of survival data, available in the literature, as well as simulated data, providing some additional hints for the adequate estimation of the EVI, high quantiles and right endpoint of $X$, whenever dealing with random censoring on heavy right tails. 
See \cite{G+N-2010, G+N-2011}, who   gave special attention to the estimation of a positive EVI, under random censoring. Under such a scheme, any EVI-estimator, the basis for the estimation of all other parameters of extreme events, needs to be slightly modified to be consistent.  
The MM estimator, in \eqref{Eq-MM}, valid for a general tail, was also considered for a comparative study. 
\vspace{.25pc}

Some authors have drawn attention to the fact that there might be practical problems with the use of unbounded Pareto distributions, for instance when there are natural upper bounds that truncate the probability tail. The authors in \cite{A+M+P-2006}  derived the ML estimator for the tail index of a truncated Pareto distribution with right truncation point $T$. The Hill EVI-estimator, in \eqref{Eq-Hill},  is then obtained from this ML EVI-estimator, letting \mbox{$T \to \infty$}. The problem of EV estimation under (right) truncation was also considered in \cite{Nuyts-2010}, who proposed a similar estimator for the tail index and considered trimming of the number of extreme OSs, and in \cite{Clark-2013}, who has also  drawn attention to possible truncation in Pareto tail modelling.
Given that in practice one does not always know whether the distribution is truncated or not,  estimators for extreme quantiles both under truncated and non-truncated Pareto-type distributions were considered in \cite{B+FA+G-2016}. 
 A truncated Pareto QQ-plot and a formal test for truncation  to help deciding between a truncated and a non-truncated case were also proposed. 
In this way, we have enlarged the possibilities of extreme value modelling using Pareto tails, offering an alternative scenario by adding a truncation point $T$ that is large with respect to the available data. In the mathematical modelling we hence let  $T\rightarrow\infty$ at different speeds comparatively to the limiting fraction $(k/n\rightarrow  0)$ of data used in  the extreme value estimation. This work was motivated using practical examples from different fields of applications, simulation results, and some asymptotic results (see also the associated preprint, \cite{B+FA+G+M-2014}).

\subsubsection{Comparison at asymptotic optimal levels}
The asymptotic comparison at optimal levels, in the sense of minimal {\it asymptotic mean square error} (AMSE), of classes of estimators of parameters of extreme events has been another topic on which I have carried out  research, together with several elements of the PORTSEA. I only mention  two international articles, in which this topic was addressed individually:
\cite{G+N-2008}, by Gomes and C. Neves, where an asymptotic comparison of the MM estimators and alternative classical EVI estimators was carried out;
\cite{G+HR-2010b}, by Gomes and Henriques-Rodrigues, with a comparison of classical estimators  that feeds the challenge for the consideration of {\it reduced-bias} (RB) estimators, also compared asymptotically at optimal levels in a few articles, such as  can be seen later in  {\bf   \ref{Sec-VIes-Reduzido}}.
  
\subsubsection{The role of generalized means in statistical EVT\label{Sec-GMs}}
Several {\it generalized means} (GMs) were considered by members of PORTSEA around the year 2000 to estimate the EVI. 
Among them, we mention the  {\it power mean of exponent $p$} (PME$_p$) of the log-excesses, introduced by Gomes and Martins  in \cite{G+M-1999},   and further studied in \cite{G+M-2001}. Those  EVI-estimators are valid for a positive EVI, and are defined by 
	\begin{equation*}
{\rm PME}_p (k)\equiv 
{\rm PME}_p(k; {\underline{X}}_n):= 
	\left(	\frac{M_{n}^{(p)}(k)}{\Gamma(p+1)}\right)^{1/p},\quad {\rm for} \quad  p>0 \quad
	[ {\rm PME}_1(k) \equiv H(k)],
	\label{Eq-PM-EVI}
\end{equation*}
with $\Gamma(\cdot)$ denoting the complete Gamma function and $M_n^{(p)}(k)$ given in \eqref{Eq-M_n^p}. 
\vspace{.25pc}

Further notice that the Hill EVI-estimators, in \eqref{Eq-Hill}, being the means of the $k$ excesses of the logarithm of the ordered observations, can be considered as geometric means, or equivalently, means of order $p = 0$, of the set of  core statistics, 
$$
U_{ik}:= X_{n-i+1:n}/X_{n-k:n}, \quad 1\leq i\leq k<n.
$$
Instead of such a geometric mean, and after the development of some preliminary publications on the subject, the authors  in \cite{B+G+P-2013, B+G+P-2014b},  Brilhante,  Gomes and D. Pestana,   considered in a more general way the {\it mean of order $p \geq 0$} (MO$_p$) of these statistics.  Note that it is further possible to  consider $p\in\mathbb{R}$ (see \cite{C+G+B+W-2016a}), and the  class of MO$_p$ EVI-estimators,
\begin{equation}
{\rm H}_p(k) :=\left\{
\begin{array}{l}
\tfrac{1}{p}\left({{1-\left(\tfrac{1}{k}\sum\limits_{i=1}^k  U_{ik}^p\right)^{-1}}}\right),~   \mbox{if}\quad  p <1/\xi,\ p\neq 0,\\
\ln \left(\prod\limits_{i=1}^k U_{ik}\right)^{1/k} =  \tfrac{1}{k}\sum\limits_{i=1}^k V_{ik}=: {\rm H}(k),~
 \mbox{if}\quad  p=0,
\end{array}
\right.
\label{Eq-MOP}
\end{equation}
with ${\rm H}({k})\equiv {\rm H}_0(k)$, the Hill EVI-estimators.
\vspace{.25pc}

In addition to some relevant articles involving bias reduction and the use of GMs, to be referred   in  {\bf  \ref{Sec-VIes-Reduzido}}, we mention now  the following ones:
\cite{G+B+C+P-2015a}, with the study of a partially RB MO$_p$ class of EVI-estimators;
 \cite{G+F-2020}, where the tail estimation is done via GMs;
\cite{P+G+C+N-2020a}, with an asymptotic comparison of some non-reduced bias estimators based on different GMs;
\cite{P+G+C+N-2020b}, where a simulation study of an EVI estimator based on the Lehmer’s mean of order-$p$ was provided (see also the PhD thesis by Helena Penalva, \cite{Penalva-2017}).

\subsection{Resampling  methodologies (RMs) and bias reduction in statistical EVT\label{Sec-Jackk+Boot}} 

\subsubsection{Generalities\label{Sec-RB-Generalities}}
Resampling methodologies like the jackknife (Quenouille, \cite{Quenouille-1956}; Tukey, \cite{Tukey-1958}),  the {\it generalized jackknife} (GJ) (see \cite{G+S-1972}) and the bootstrap (Efron, \cite{Efron-1979})  have been used in statistical  EVT for the estimation of parameters of extreme events, among which I mention  the EVI, in \eqref{Eq-EVI-xi},  and the EI, in \eqref{Eq-EI}.
These methodologies have frequently answered positively to the question whether the combination of information can improve the quality of estimators of parameters/functionals, a discussion that can be seen in \cite{MIG-1995a}, a paper associated with a presentation at the ``{II \it Congresso Anual da SPE}’’ in 1994,  and that initiated the study of {\it reduced-bias} (RB) estimators of parameters of extreme events and of threshold selection in statistical EVT through resampling techniques. 
I believe that my presentation   at the II SPE  meeting  was responsible for the co-supervision, together with Manuela Neves, of a new PhD student,  Maria Jo\~{a}o Martins, who worked tirelessly on this topic and related ones, essentially in the field of  semi-parametric RB estimation of parameters of rare events, having completed her thesis in 2001 (see \cite{MartinsMJ-2001}). 
 The use of the jackknife methodology in  a semi-parametric framework of {\it Statistics of Extremes}   was initiated in the sequel of an invitation to give a seminar at Erasmus University of Rotterdam in 1998.
Apart from the initial 1995 paper, I  thus further  mention two individual  articles  that I published in this area (see \cite{MIG-1999b, MIG-1999c}). 
 I cannot also  fail to mention Fernanda Figueiredo, with a PhD completed in 2003 (see \cite{Figueiredo-2003}), in Statistical Quality Control, but who was partially `{\it converted}’ to this area of {\it extremes},  Frederico Caeiro, with a PhD completed in 2006 (see \cite{Caeiro-2006}),  
Lígia Henriques-Rodrigues, with a PhD got in 2009 (see \cite{HenriquesR-2009}), 
and more recently Ivanilda Cabral, who completed her PhD in 2021 (see \cite{Cabral-2021}).
Generally, the use of resampling methods leads to a reliable estimation of any parameter of extreme events, as can be seen in \cite{MIG-2014b, MIG-2023c, G+F+M+N-2015d}, among others.

\subsubsection{The jackknife methodology and other initial RB procedures\label{Sec-Initial-RB+Jackknife}}
Bias reduction, and the possible use of jackknife and GJ methodologies, has been one of the fields of {\it statistical} EVT in which I have invested the most in recent years, with the collaboration of several colleagues and Doc and Post-Doc students.
\vspace{.25pc}

The main objectives of the {\it jackknife methodology} are the estimation of  {\it bias} and {\it variance}  of a statistic, only by
manipulation of observed data $\underline x$, and the 
 {\it building} of estimators with  bias and {\it mean squared error} (MSE)  smaller than those of an initial set of estimators.
Let $\underline X=(X_1,\dots,X_n)$ be a sample from an
underlying model $F$, and let $T_n=T_n(\underline X,F)$ be
an estimator of a functional $\theta(F)$, or eventually of a
parameter $\theta$, in the case $F$ is known up to unknown
parameters. 
{\it Quenouille-Tukey's method} works in the following way: Let $T_{n,i}$, $i =1, 2,
\dots, n$,  be an estimator (with the same
functional expression of $T_n$) but based on the subsample of
size $n-1$, obtained from the original sample by removing its
$i$-th element, $1\leq  i\leq n$. Let us build
$$\widetilde T_{n,i}  = nT_n  -
(n-1)T_{n,i}, \quad i = 1,\dots, n,$$  
the so-called {\it pseudo-values of the jackknife}. 
The {\it pure jackknife estimator}, 
$$
T_n^J=\frac{1}{n}\sum_{i=1}^n\widetilde T_{n,i} =
nT_n-(n-1)\overline T_{n,i},\qquad  \overline T_{n,i}=\frac{1}{n}\sum\limits_{i=1}^n T_{n,i},
$$
enables us to eliminate the term of order $1/n$ of a bias of
the form $E(T_n)=\theta+a_1(\theta)/n+o(1/n)$.
More generally, if we have access to two estimators
$T_n^{(1)}$ and $T_n^{(2)}$ of $\theta$, where the second
estimator may obviously be built on the basis of the first
estimator and Quenouille-Tukey method,   if we get to know asymptotic 
expansions for the biases of those estimators and we may
compute
$\alpha_n=\frac{E[T_n^{(1)}-\theta]}{E[T_n^{(2)}-\theta]}$,
then the GJ estimator,
$$ 
T_n^G=\frac{T_n^{(1)}-\alpha_n T_n^{(2)}}{1-\alpha_n}
$$
is unbiased for the estimation of $\theta$. Under specific
regularity conditions, the MSE of $T_n^G$ can be 
smaller than that of either $T_n^{(1)}$ or $T_n^{(2)}$, but the variance increases.
Articles on this topic and of great interest are co-authored with Maria Jo\~{a}o Martins and Manuela Neves (see \cite{G+M+N-2000, G+M+N-2001, G+M+N-2002b}), 
 all using the jackknife methodology to estimate a positive EVI, with   different alternatives to  the Hill estimator, in \eqref{Eq-Hill}. And indeed, most of the GJ EVI-estimators are linear combinations of Hill estimators at different levels. 
\vspace{.25pc}

In \cite{G+M-2002}, again in a context of regularly varying tails, particular but interesting cases of the  {\it maximum likelihood} (ML) and {\it least squares} (LS) estimators proposed by   Feuerverger and P Hall (see \cite{F+H-1999}) were analysed by Gomes and Martins. 
These estimators are alternatives to the Hill EVI-estimators, and have   in mind a reduction in bias   without increasing MSE,  provided we may use EV  data relatively deep into the sample. 
A simple external estimation of a generalized shape second order parameter  was  considered, to build an ML estimator of $\xi$, with a dominant component of asymptotic bias of smaller order than that of the Hill estimator.
And this was the starting point for most of the developments achieved after 2005, to be mentioned in {\bf \ref{Sec-VIes-Reduzido}}.
See also, \cite{C+G-2015b}, an article by Caeiro and Gomes, where an ML estimator of a positive EVI was revisited. 
I further mention some of the initial articles on RB estimation of extreme events’ parameters, usually under the framework of a regularly varying right-tail: 
\cite{G+M-2001, G+M-2004, C+G-2002a, C+G-2002b, G+C+F-2004a, G+F+M-2005a}. 
 In most of these articles, the reduction of bias was done at the cost of a significant increase in variance, as was already usually the case when using the Jackknife methodology at an initial stage, something considered as the `{\it trade-off between bias and variance}’.
\vspace{.25pc}

 In \cite{G+M-2003}, Gomes and Miranda introduced   an RB EI-estimator, on the basis of the use of the GJ methodology. 
I further mention another relevant article on this same topic, in which I am co-author, together with A. Hall and Miranda (see \cite{G+H+M-2008d}), where  the GJ methodology was used together with subsampling techniques, enabling the  improvement of an EI-estimation.
\vspace{.25pc}

In \cite{G+P+M-2005b}, again in a context of regularly varying tails, a generalization of the classical Hill estimator of a positive EVI was analysed. The members of this general class of estimators were not asymptotically more efficient than the original one, and   GJ EVI-estimators based on two members of such a class were then proposed. 
Other GJ EVI-estimators have been considered and studied in   \cite{G+M+V-2007b}. 
\vspace{.25pc}

The importance of the GJ methodology in the construction of a reliable semi-parametric estimate of any parameter of extreme or even rare events was later revisited by Gomes, Martins and M Neves, in \cite{G+M+N-2013g}.

\subsubsection{Research mainly involving the bootstrap\label{Sec-Bootstrap-only}}

The main objective of statistical EVT is the prediction of rare events, and its primary problem has been the estimation of the EVI, $\xi$, usually performed on the basis of the largest $k$ OSs in the sample or on the excesses over a high level $u$. 
The question that has been often addressed in practical applications of  EVT is the choice of either $k$ or $u$, and an adaptive estimation of $\xi$. 
The Bootstrap methodology has been used successfully in {\it statistics of extremes}, and   we have had the collaboration of several members of our School.
The bootstrap  has  indeed proven to be widely useful in estimating the {\it optimal sample fraction} (OSF). And among other articles published nationally and internationally, I first mention the  seminal paper by 
Gomes \& O Oliveira  (see \cite{G+O-2001}), who were mainly interested in the use of the bootstrap methodology to estimate the EVI adaptively, and although the methods provided might be applied, with adequate modifications, to the general domain of attraction of $G_\xi, \xi\in \mathbb{R}$, in \eqref{Eq-EVI-xi}, only heavy tails (i.e.\ $\xi>0$) were considered. 
Indeed, the robustness of Hall's bootstrap methodology, in \cite{HallP-1990}, to the choice of sub-sample sizes, led   to the computational  study of the performance of a bootstrap estimator of the OSF and the corresponding adaptive estimator of $\xi$, of the type of the ones studied in \cite{D+H+P+TP-1999, D+H+P+V-2001}, but with the use of a  simpler auxiliary statistic converging to zero, which is merely the difference of two estimators with the same functional form as the estimator under study, computed at two different levels, like $k$ and $\lfloor k/2\rfloor$. These bootstrap methodologies were compared, by Monte Carlo simulation,  with other methodologies available in the literature like the ones proposed in \cite{B+V+T-1996a, B+V+T-1996b, D+K-1998}, together with the proposal in  \cite{H+W-1985}.
A few problems related to the use of the bootstrap methodology based on  spacings of OSs,  with the construction of new tail index estimators,  were put forward in \cite{N+M+G-2001},  where new estimators of the tail index were obtained and studied and a bootstrap estimator of the MSE was also considered. 
\vspace{.25pc}

The class of MO$_p$ EVI-estimators, already defined in \eqref{Eq-MOP}, depends on the extra tuning parameter $p\geq 0$ (or possibly $p\in\mathbb{R}$), which makes it very flexible, and even able to overpass most of the `classical' and even  RB EVI-estimators. Apart from a simulation study that reflects such an assertion, the authors in \cite{B+G+P-2012a}, Brilhante, Gomes and D.\ Pestana,   advanced with a fully non-parametric double bootstrap algorithm for the choice of $p$ and $k$, the number of upper OSs used in the estimation,  and further provided applications of the algorithm to both simulated and real data. 
 \vspace{.25pc}

The authors in \cite{G+F+N-2012b, G+F+N-2013c}, Gomes, F. Figueiredo and M. Neves,  discussed  different algorithms for the adaptive estimation of a positive EVI, $\xi$. Apart from classical   EVI-estimators,  the consideration of associated second-order RB  EVI-estimators was suggested, and   the use of bootstrap computer-intensive methods for the adaptive choice of thresholds was put forward.  
 \vspace{.25pc}

Caeiro and Gomes, in \cite{C+G-2013d}, dealt with the estimation, under a semi-parametric framework, of a negative EVI.
Apart from the usual integer parameter $k$, related to the number of upper OSs involved in the estimation, the estimator depends on an extra tuning parameter, which makes it highly flexible and possibly second-order unbiased for a large variety of models. Both parameters were chosen via bootstrap methodology.
In \cite{C+G-2014c}, the adaptive estimation of either $k$ or $u$ through a nonparametric bootstrap methodology was considered. An improved version of Hall’s bootstrap methodology was introduced and compared with   the double bootstrap methodology. 
 \vspace{.25pc}

An overview of the bootstrap methodology together with its possible use in the reliable estimation of any parameter of extreme or even rare events was provided by Gomes, Caeiro, Henriques-Rodrigues and Manjunath,  in \cite{G+C+HR+M-2016b}.

\subsubsection{Semi-parametric estimation of second-order  parameters\label{SubSec-Ordem2}} 

The role of our School in this area has also been important, particularly due to the emphasis that has been given in our working group to the development of second-order   {\it minimum-variance reduced-bias} (MVRB) estimators of several parameters of extreme events, just as will be seen in   {\bf   \ref{Sec-VIes-Reduzido}}.
For heavy right-tailed models, i.e.\ models in ${\cal D_M}^+$, which necessarily have a regularly varying right tail, such that 
$$1-F(x)=C x^{-1/\xi}\mathbb{L}(x), 
$$
with $\mathbb{L}(\cdot)$ a slowly varying function, i.e.\  a positive measurable function such that  
$$
\lim\limits_{t\rightarrow\infty} \mathbb{L}(tx)/\mathbb{L}(t)=1, \mbox{ for all }x>0, 
$$
we can pay attention to the speed of convergence in the previous limiting relation, and to introduce a sub-class of ${\cal D_M}^+:= {\cal D_M}(G_{\xi>0})$, since this rate of convergence, which also controls the convergence speed of the sequence of maxima, linearly normalized, to the GEV  CDF, in \eqref{Eq-EVI-xi},  depends upon a function  $A(t)=\xi\beta t^\rho$, where  $(\beta, \rho)\in \mathbb{R}\setminus\{0\}\times \mathbb{R}^{-}$ is a vector of second-order parameters, being $F$ a model in Hall-Welsh class of models. With the usual notations, $F^\leftarrow(y):= \inf\{x:F(x)\geq y\}, 0\leq y\leq 1,$ for the {\it generalized inverse function}  of $F$, and 
$
U(t):= F^\leftarrow (1-1/t), \ t\geq 1,
$
for the {\it reciprocal tail quantile function}, 
\begin{equation}
U(t)= C t^\xi \left(1+A(t)/\rho +o(t^\rho)\right),\quad A(t)=\xi\beta t^\rho,\ \xi>0, \ \beta\neq 0, \ \rho<0.
\label{Eq-Hall-Welsh}
\end{equation} 

The estimation of the  “shape” second order parameter, $\rho<0$, in \eqref{Eq-Hall-Welsh}, was addressed in \cite{MIG-2000}, as well as in \cite{FA+G+H+L-2001, G+H+P-2002a, FA+G+H-2003, G+HR+P+P-2010b, C+G-2014a, HR+G+FA+N-2014}.
   For an asymptotic comparison of two classes of estimators of the ``shape’’ second-order  parameter see \cite{C+C+G-2021}.
 \vspace{.25pc}
  
The “scale” second-order  parameter $\beta$, also in \eqref{Eq-Hall-Welsh}, was estimated, sometimes in a non-isolated way, in \cite{G+M-2002, C+G-2006, C+G-2012, G+HR+P+P-2008e}. 
 As a generic article on the subject, written by Fraga Alves, Gomes and de Haan,
%  and in ${\cal D_M}(G_\xi)$,         $\xi\in\mathbb{R}$, 
  I mention \cite{FA+G+H+N-2007a}.
  \vspace{.25pc}
 
The need to develop techniques for the adaptive  choice of the optimal level for RB estimators of   the tail index and of other parameters of extreme events, leads me to believe that we still need to advance on this topic, considering  conditions of a higher order than the second one.
A good contribution was initially done in \cite{C+G+HR-2009, G+M+P-2009a}, and in a few other articles to be included in  {\bf Section \ref{SubSec-PORT}}.
 In \cite{C+G-2015a}  further progresses were achieved, with the reduction of bias in the estimation of the ``shape’’ second-order  parameter.

\subsubsection{Advances in bias reduction\label{Sec-VIes-Reduzido}} 
 From 2005 onwards, the adequate estimation of second order parameters, essentially due to developments achieved in articles   referred to in {\bf \ref{SubSec-Ordem2}}, allowed us to maintain the variance and eliminate the dominant component of asymptotic bias, drastically improving the behaviour of the estimators for all $k$.
Considering only the EVI-estimation, and essentially for heavy tails, details about these new MVRB estimation methods can be seen in: 
\cite{C+G+P-2005, G+P-2007a, G+M+N-2007a, G+HR-2008}, with the accommodation of bias  performed in the excesses over a high level; 
 \cite{G+H+HR-2008c}, with bias accommodation performed in the weighted excesses of the top log-observations;
 \cite{G+HR+V+V-2008f};
 \cite{C+G+HR-2009}, under a third-order framework; 
 \cite{C+G-2010,  C+G-2011b, C+G-2014b, G+C-2014}; 
 \cite{G+B+C+P-2015a}  and \cite{C+G+B+W-2016a}, both for GMs and already referred in {\bf \ref{Sec-GMs}};
 \cite{C+C+G-2016, G+HR-2016, G+HR-2017}, with comparisons of a large diversity of competitive estimators; 
 \cite{G+B+P-2016a};
 \cite{G+P+C+N-2016d}, in which we draw attention to the need to debate the topic of `efficiency vs robustness’ in {\it statistics of extremes};
\cite{C+C+G-2018, C+G+HR+C-2020}.
 \vspace{.25pc}
 
 The most simple MVRB EVI-estimators are the ones in \cite{C+G+P-2005}. In Hall-Welsh class of models, in \eqref{Eq-Hall-Welsh}, the dominant component of bias of the Hill estimators H$(k)$, in \eqref{Eq-Hill}, given by $A(n/k)/(1-\rho)$ (see \cite{Haan+P-1998}),  is estimated at a large adequate threshold and directly removed from the Hill EVI-estimators. The functional expression of these MVRB EVI-estimators is thus given by
\begin{equation*}
\overline {\rm H}(k) = \overline {\rm H}_{\hat\beta, \hat\rho}(k) :=  {\rm H}(k)\big(1-\hat\beta (n/k)^{\hat\rho}/ (1-\hat\rho) \big),
\label{Eq-Hill-Bias}
\end{equation*}
with ${\rm H}(k)$ the Hill  EVI-estimators, in \eqref{Eq-Hill}, and where $\hat\beta$ and  $\hat\rho$ need to be  adequate consistent estimators of  the generalized scale and shape  second-order parameters $\beta$ and $\rho$.
\vspace{.25pc}

With the inclusion of VaR estimation, mainly improving the performance of classical estimators for high quantiles (see \cite{Weissman-1978}, a pioneer paper in the topic, by Ishay Weissman), as well as the estimation of additional parameters of rare events, I mention:
\cite{G+F-2006, G+P-2007b, G+P-2009b, B+F+G+V-2008a, C+G-2008, C+G-2009, G+C+F+HR+P-2020a, G+C+F+HR+P-2020b}.
 Adaptive `threshold’ choices in RB estimation can be seen in \cite{G+M+P-2009a}, among others.
The authors in \cite{G+M+P-2011b}  considered second-order MVRB estimators of a positive EVI,  and associated estimation of the  VaR at a level $q$. For those MVRB estimators,  the use of bootstrap computer-intensive methods for the adaptive choice of thresholds was proposed. 
 \vspace{.25pc}

Asymptotic comparisons at optimal levels of tail index estimators, with the inclusion or challenges for the consideration of RB estimators, can be seen in \cite{G+P+C-2009b, G+HR-2010b, C+G-2011a,  C+G-2013b, C+C+G-2022}.
\vspace{.25pc}

The problem of bias reduction was also addressed in \cite{HR+G-2022}, by Henriques-Rodrigues and Gomes, where special emphasis was given to the role of Box-Cox transformations in the reduction of bias. 
\vspace{.25pc}

For a critical review of RB estimators of parameters of rare events, see \cite{MIG-2007b}, Chapter~6 of the third edition  of the book by Reiss \& Thomas (see \cite{R+T-1997}). See also the more recent and already mentioned critical reviews, \cite{B+C+G-2012, G+G-2015}.

\subsection{The PORT methodology\label{SubSec-PORT}}

The acronym PORT  was  introduced in \cite{AS+FA+G-2006}.
The idea is very simple: In the case of shifts in the data $X_j$, defined by $Z_j:= X_j+\lambda$, $1\leq j \leq n$, for any $\lambda\in \mathbb{R}$,
the aim is to find semi-parametric EVI-estimators   that are invariant to changes in  the location (as well as in the scale), as happens with the EVI itself.
These estimators are based not directly on the original sample, $\underline{\bf X}_n:=\left\{ X_j\right\}_{j=1}^n$, but rather on the sample of excesses above an empirical  quantile, $X_{n_s:n}:=X_{\lfloor ns\rfloor+1:n}$, $0<s<1$, i.e.\ on
\begin{equation}
\underline{\bf X}_n^{(s)}:= \left\{X_{n-j+1:n}-X_{n_s:n},~ 1\leq j \leq n-n_s \right\}, \quad n_s:= \lfloor ns\rfloor+1,
\label{Eq-Excessos-quantil}
\end{equation}   
where $\lfloor x\rfloor$ denotes the integer part of $x$. 
For underlying models $F(\cdot)$ with a finite left endpoint, one can even consider $s=0$, that is, excesses above the sample minimum.
Note that the classical EVI-estimators, such as the Hill estimator, in \eqref{Eq-Hill},   the M estimator, in \eqref{Eq-Momentos-DEH},  and the MM estimator, in \eqref{Eq-MM},  are invariant to  changes in the scale, but depend strongly   on changes in the location, making it wise to consider the PORT-Hill, PORT-M and PORT-MM EVI-estimators, 
\begin{equation}
{\rm H}_s(k) := {\rm  H}(k; \underline{\bf X}_n^{(s)} ), \quad {\rm M}_s(k) := {\rm  M}(k; \underline{\bf X}_n^{(s)} ),\quad {\rm MM}_s(k) := {\rm  MM}(k; \underline{\bf X}_n^{(s)}),
\label{PORT-H+M+MM}
\end{equation}
based on the sample of excesses in \eqref{Eq-Excessos-quantil}.
The \mbox{${\rm H}_s\equiv$ PORT-H} and \mbox{${\rm M}_s\equiv$PORT-M}  EVI-estimators, in \eqref{PORT-H+M+MM}, were initially studied in \cite{G+FA+AS-2008b}. 
The \mbox{${\rm MM}_s\equiv$ PORT-MM}   EVI-estimators, also in \eqref{PORT-H+M+MM},
 were studied in \cite{FA+G+H+N-2009}, already mentioned.
\vspace{.25pc}

The PhD thesis by Ligia Henriques-Rodrigues (see \cite{HenriquesR-2009}), discussed at FCUL and under my supervision, was essentially in the topic of this subsection, but with a particular focus on RB estimation.
The PORT-MVRB EVI-estimators were studied in 
\cite{G+HR+M-2010a, G+HR+M-2011a, G+HR+FA+M-2013f, HR+G-2018}.
The PORT methodology was also applied to PWM estimation, in 
\cite{G+C+HR-2012a, C+G+HR-2016b}, 
to {\it best linear unbiased estimators} (BLUE), in \cite{HR+G-2014},
and to GMs, in  \cite{G+HR+M-2016c}, where a more reliable EVI-estimation was achieved. 
For a PORT estimation of the second-order parameters, see \cite{HR+G-2013, HR+G+M-2015a}.
 \vspace{.25pc}
 
For the estimation of other parameters of extreme events, such as the VaR, PORT estimators are based on the properties of those parameters, given changes in location (see the pioneer paper, \cite{AS+FA+G-2006}, among other articles on the topic, such as \cite{HR+G-2009, F+G+HR+M-2012} and \cite{F+G+HR-2017}, where GMs are additionally considered).
 \vspace{.25pc}

I further refer other articles (see \cite{G+HR-2010a,  G+HR-2011,  G+HR-2013}),  in which we advanced with methods for the choice of the OSF in PORT RB estimation.

\subsection{Extremes under non-regular frameworks\label{Sec-Non-Regular}}
There are a great diversity of situations in {\it statistics of extremes}  where  it is not possible to obtain a normal asymptotic behaviour of some estimators of parameters of rare events. In the context of the specific GMs, in \eqref{Eq-MOP},  and for the EVI-estimation, the authors in \cite{G+HR+P-2021, G+HR+P-2022a, G+HR+P-2022b}  advanced with studies of the non-normal asymptotic behaviour of estimators in a context of non-regularity, and with the possibility of an almost degenerate estimation of the EVI, having still left much research open.

\subsection{Aplications\label{Sec-Aplicacoes}}

\subsubsection{Extremes and Environment}

It is sensible to mention the editions of two books on environmental themes, \cite{FA+G-2006, FA+G+H+T-2007b}. 
\vspace{.25pc}

Among many other   articles involving this topic, I mention the following ones: 
\cite{G+P-1986a}, with the introduction of non-classical EV models and their application to climatological data;
\cite{MIG-1993b},   with developments in the estimation of parameters of rare events in environmental time series.
And there are some additional international articles on this topic:
\cite{N+G+FA-2011} advanced with an assessment of the extreme limits of nitriding in aluminium extrusion;
\cite{G+HR+C-2013e}  outlined a refined method for estimating a light right tail, with applications to environmental data;
\cite{G+HR+F-2015e}  developed different methods for estimating parameters of extreme events based on resampling methods  and applied those methods to environmental and financial data;  
\cite{L+F+G+L-2016}  applied Birnbaum-Saunders regression models  to environmental data (for a review of this type of models, applied to various data related to natural disasters, see \cite{L+L+G+F-2019}).
 \vspace{.25pc}
 
 At a national level, I also mention \cite{G+P-2019}, who addressed the justification for the relevance of statistical EVT in predicting earthquakes.

\subsubsection{Extremes in Finance, Insurance and Telecommunications}
My work in this field of applications dates   to 1981 (see \cite{MIG-1981b}), where several extreme models in {\it Economics} were discussed.
 I also mention \cite{MIG-2004a}, on extremes and risk management in finance,
\cite{MIG-2004b}, on the generic role of stochastic processes in telecommunications traffic, and\cite{G+P-1986b}, with the introduction of non-classical EV models and their application to actuarial data.
I also cannot fail to mention the publication of the book,\cite{FA+G+H+N-2011}, on the role of risk analysis and statistical EVT in Insurance and Finance.

\subsubsection{Extremes in Athletics} 
 The importance of light tails in {\it Sports}, with the estimation of several useful parameters, was highlighted in \cite{G+P-2009a}, 
  and in \cite{HR+G+P-2011}.
The Birnbaum-Saunders model was also applied to athletics data, in \cite{G+F+L-2013b}.
The generic role of statistical EVT in {\it Sports} was emphasized in \cite{HR+G+P-2015b}.

\subsubsection{Extremes in Biometry}

With applications to biometric data, we were able to count only three international articles, \cite{G+N-2010, G+N-2011, G+F+L-2013a}.

\subsubsection{Extremes in Statistical Quality Control}

On this topic, and following Fernanda Figueiredo's PhD thesis, already referred in {\bf \ref{Sec-RB-Generalities}}, I first mention  some articles essentially in the area of {\it Statistical Quality Control} (SQC), but in which various methods are used involving {\it Order Statistics and Extremes}: \cite{F+G-2004}, with details on the role of moving maxima and the total median in SQC;
 \cite{F+G-2013, F+G-2015}, on the role of asymmetric-normal distributions in SQC and in eliminating risk in modelling real data;
\cite{F+G-2016}, on the role of the total median in monitoring contaminated normal data, subject to various types of contamination;
\cite{F+F+G-2018}, with the development of acceptance sampling plans to reduce the risk associated with chemical compounds.
\vspace{.25pc}

And I again recall the need to move forward with applications to real data, regarding  the progresses made in the use of results on pre-asymptotic behaviour in EVT, in the study of the reliability of   high-dimensional coherent systems, a topic already  referred to in {\bf \ref{Sec-Velocidade}}.

\section{A few additional  topics and final comments\label{Sec-Final}} 
I also mention applications of {\it extremes} to {\it queuing theory},  in \cite{MIG-1980b}, where basic facts in EVT were also reviewed.
I further would like to refer a software package (see \cite{M+C+G+FA-2013}), on  MO$_p$ PORT Hill and high quantile estimates, essentially under the responsability of Bangalore Manjunath, a PosDoc student of Isabel Fraga Alves and I, who got his PhD in Siegen, Germany, under the supervision of Rolf-Dieter Reiss, and who  is currently a Professor at the {\it School of Mathematics and Statistics}, University of Hyderabad.
And our School also has strong groups in the general area of {\it extremes and risk modelling}, where some of the previously mentioned articles are  located, together with further research  that will be next mentioned. 
In addition to the edition of a book,  \cite{G+H+P+CC+FA-2003}, dedicated to the areas of statistical EVT, risk and resampling techniques, I refer to the following article: \cite{B+G+P-2019}, on risk modelling of extreme events in generalized Verhulst models. 
\vspace{.25pc}

I also would  like to mention: a discussion article, \cite{MIG-1990a},   about a paper by AC Davison and RL Smith, related to the POT methodology and published in the {\it J. Royal Statistical Society B}; 
an obituary of J Tiago de Oliveira (see \cite{MIG-1994a}), published in the {\it J. Royal Statistical Society A}; 
an additional editorial note, \cite{G+AT-2017},   congratulating Sir David Cox for an ISI International Prize of Statistics;
and two critical reviews, \cite{MIG-1999a}, on the first edition of the book by RD Reiss \&  M Thomas, and
\cite{MIG-2009}, on a book by Natalia Markovich, about research and practice in non-parametric analysis of heavy-tailed univariate data.
\vspace{.25pc}

The challenges and opportunities in this area are still enormous, as can be seen in \cite{MIG-2022a}, written in Portuguese and recently published in {\it Mem\'{o}rias da Academia das Ci\^{e}ncias de Lisboa}.
\vspace{.25pc}

Just as has been enhanced above, the fields of EVT, in which  PORTSEA's contribution has been important, are very diverse.
In addition to a vast group with innovative work in the area of Parametric, Semi-parametric and Non-parametric Estimation of Parameters of Extreme Events, PORTSEA has strong groups in the areas of\begin{itemize}
\vspace{-.5pc}
\item [---] Statistical Choice of Extremal Models, 
\vspace{-.5pc}
\item  [---] Extremes and Risk Modelling, 
\vspace{-.5pc}
\item [---] Environmental Extremes, 
\vspace{-.5pc}
\item [---] Extremes of Dynamical Systems, 
\vspace{-.5pc}
\item [---] Extremes of Dependent Sequences, and
\vspace{-.5pc}
\item [---] Spatial Extremes. 
\end{itemize}
And, as a prediction, I hope that  we shall soon have   a group  of
\begin{itemize}
\vspace{-.5pc}
\item [---] Extremes in Genetics and another in 
\vspace{-.5pc}
\item [---] Extremes in Epidemic Situations.
\end{itemize}

\noindent
In view of the results obtained, I am led to believe that our ‘{\it Escola de Extremos}’  (or our PORTSEA) has in fact achieved a healthy growth in the area. The dynamism of the Group has allowed a high international recognition of the School of Extremes in Portugal, a country of `{\it nice extremists}' at one end of Europe.

\subsection*{Acknowledgements} 
This work is partially financed by national funds through FCT---Funda\c{c}\~{a}o para a Ci\-\^{e}n\-cia e a Tecnologia under the project \href{https://doi.org/10.54499/UIDB/00006/2020}{UIDB/00006/2020}, and by HiTEc Cost Action CA21163. I also would like to thank the reviewers for their comments and suggestions.

%%% REFERENCES %%%

\EditInfo{February 23, 2024}{April 29, 2024}{Ana Cristina Moreira Freitas, Diogo Oliveira e Silva, Ivan Kaygorodov, and Carlos Florentino}

\end{document}